\documentclass[fleqn,usenatbib]{mnras}

\usepackage{newtxtext,newtxmath}
\usepackage[T1]{fontenc}
\usepackage{ae,aecompl}

\usepackage{graphicx}	% Including figure files
\usepackage{amsmath}	% Advanced maths commands
\usepackage{amssymb}	% Extra maths symbols
\usepackage{verbatim}
\usepackage{sansmath}

\newcommand{\de}{\rmn{d}}
\newcommand{\Msun}{M_{\odot}}
\newcommand{\kmsec}{\rmn{km}\rmn{s}^{-1}}
\newcommand{\kpc}{\rmn{kpc}}
\newcommand{\pc}{\rmn{pc}} 
\newcommand{\Msunppcc}{\Msun\pc^{-3}} 
\newcommand{\mas}{\rmn{mas}}
\newcommand{\masyr}{\rmn{mas}~\rmn{yr}^{-1}}
\newcommand{\Sect}[1]{Section~\ref{#1}}
\newcommand{\Eq}[1]{equation~(\ref{#1})}
\newcommand{\Fig}[1]{Fig.~\ref{#1}}
\newcommand{\bldPsi}{\mathbf{\Psi}}
\newcommand{\bldpsi}{\bpsi}

\newcommand{\tarcsec}{\rmn{arcsec}}
\newcommand{\MVobs}{M_V^\rmn{obs.}}
\newcommand{\bd}{\bmath{d}}
\newcommand{\bx}{\bmath{x}}
\newcommand{\bv}{\bmath{v}}
\newcommand{\diprime}{\bmath{d}_i'}

\newcommand{\changes}[1]{#1}%{\textbf{#1}}
\newcommand{\newchanges}[1]{#1}%{\textbf{#1}}

\title[The local dynamical matter density]{The dynamical matter density in the solar neighbourhood inferred from Gaia DR1}

\author[A. Widmark and G. Monari]{
Axel Widmark,$^1$\thanks{E-mail: axel.widmark@fysik.su.se} 
and Giacomo Monari$^{1,2}$ 
\\
$^1$The Oskar Klein Centre for Cosmoparticle Physics, Department of
Physics, Stockholm University, AlbaNova, 10691 Stockholm, Sweden\\
$^2$Lebniz Institut f\"{u}r Astrophysik Potsdam, An der Sternwarte 16, 14482 Potsdam, Germany
}

\date{Accepted XXX. Received YYY; in original form ZZZ}

\pubyear{2017}

\begin{document}
\label{firstpage}
\pagerange{\pageref{firstpage}--\pageref{lastpage}}
\maketitle
% * <giacomo.monari@fysik.su.se> 2017-06-20T09:31:20.985Z:
\begin{abstract}
We determine the total dynamical density in the solar neighbourhood using the Tycho-Gaia Astrometric Solution (TGAS) catalogue. Astrometric measurements of proper motion and parallax of stars inform us of both the stellar number density distribution and the velocity distribution of stars close to the plane. Assuming equilibrium, these distributions are interrelated through the local dynamical density. For the first time, we do a full joint fit of the velocity and stellar number density distributions while accounting for the astrometric error of \changes{individual} stars, in the framework of Bayesian hierarchical modelling. We use a sample of stars whose distance extends to approximately 160 pc from the Sun. We find a local matter density of $\rho_0 = 0.119^{+0.015}_{-0.012}~\Msun\pc^{-3}$, where the result is presented as the median to the posterior distribution, plus/minus its 16th and 84th percentiles. We find the Sun's position above the Galactic plane to be $z_\odot = 15.29^{+2.24}_{-2.16}~\pc$, and the Sun's velocity perpendicular to the Galactic plane to be $W_\odot = 7.19^{+0.18}_{-0.18}~\kmsec.$
\end{abstract}


%%%%%%%%%%%%%%%%%%%%%%%%%%%%%%%%%%%%%%%%%%%%%%%%%%

%%%%%%%%%%%%%%%%% BODY OF PAPER %%%%%%%%%%%%%%%%%%

\section{Introduction}

The first measurements of the dynamical mass density in the solar neighborhood date back to the pioneering works of the Dutch astrophysicists \cite{Kapteyn1922} and \cite{Oort1932}.  These authors were the first to relate the vertical distribution and kinematics of stars near the Sun to the dynamical mass density necessary to keep the Galaxy stationary. Considering that they used methods not yet fully developed and that they had few data points of poor quality, it is remarkable that the measurements of Kapteyn and Oort are so similar to recent results \cite[see e.g. Fig.~2 in][]{Read2014}.\footnote{Kapteyn and Oort also noted a discrepancy between the dynamical mass density they measured and the one computed counting the stars in the solar neighbourhood. Kapteyn even used the term {\it dark matter} \citep[previously used only by][]{Poincare1906} to indicate this discrepancy. \newchanges{The dark matter was thought to be formed by non-emitting or faint standard matter \citep[e.g. gas, dust, dark stars, etc.,][]{KapteynLegacy}.} However, the discrepancy they observed was based on inaccurate estimates of the visible matter of the time.} 

We had to wait until the late 80s before \cite{KuijkenGilmore1989a,KuijkenGilmore1989b,KuijkenGilmore1989c,KuijkenGilmore1991} had developed methods and high quality data for the first reliable estimates of the local dynamical density, which was found to be $\rho_0=0.1~\Msunppcc$, compatible with the one observed in stars and gas.

Since the work of Kuijken \& Gilmore, due to better data and more refined methods, there have been multiple attempts to determine the local dynamical density and the integrated surface density $\Sigma_0$ of the Milky Way. An account of these results is found in \cite{Read2014}. In the late 90s, the data releases of the Hipparcos mission \citep{Hipparcos} represented a giant leap in the understanding of the structure and dynamics of our Galaxy, especially in the solar neighbourhood.  Hipparcos, being an astrometric mission, contains information about stars' position on the sky, parallax, and proper motion, extending to a distance of about $\sim 200~\pc$ from the Sun. In order to build dynamical models from this data, one had to use it either in combination with spectroscopic ground-based follow ups that would provide the missing line-of-sight velocity, or make assumptions that would allow construction from astrometric data only. This second approach was followed in studies of the local dynamical density, by \cite{Creze1998} and \cite{HolmbergFlynn2000}. While Cr\'ez\'e~et~al. measured a rather low dynamical density ($\rho_0=0.076 \pm 0.015\Msunppcc$), Holmberg~\&~Flynn found a density more in line with other studies ($\rho_0=0.102 \pm 0.010\Msunppcc$).  According to Holmberg~\&~Flynn this difference was due to the different potential used for the fit.

The next leap in understanding of the dynamics and formation of the Milky Way is expected with the Gaia mission \citep{Gaia}, which is currently collecting  astrometric data for more than a billion stars in the Milky Way, and spectroscopic data for a large subset of them. Gaia will also be complemented by ground-based spectroscopic surveys, like WEAVE \citep{WEAVE} and  4MOST \citep{4MOST}, especially for the stars too faint for having spectroscopic information from the Gaia instruments. As of 2017, only  the first Gaia data is available \citep{GaiaDR1}, containing position on the sky and $G$-band magnitudes  for stars observed a sufficient number of times. A subset of these stars  was cross-matched with the Hipparcos \citep{Hipparcos} and Tycho-2 \citep{Tycho2} catalogues, to obtain a new astrometric catalogue, called the Tycho-Gaia Astrometric Solution \citep[TGAS][]{TGAS,TGASrel}. The TGAS catalogue provides astrometric information for more than 2 million sources down to magnitude  $\sim 11.5$, with position uncertainty  $\lesssim 0.6~\mas$, parallax uncertainty $\lesssim 0.65~\mas$, and proper motion uncertainty $\lesssim 2.7~\masyr$, for 90 percent of its stars. However, the TGAS catalogue do not constitute a complete set of stars down to magnitude $\sim 11.5$. The stars in TGAS are distributed according to a complicated selection function, partly as an inheritance of the Gaia scanning law. To have a complete set of stars down to a certain magnitude we will have to wait the second data release of Gaia in 2018.

In this work, we repeat the measurement of the local dynamical density using the TGAS catalogue and the method previously applied by Cr\'ez\'e~et~al. and Holmberg~\&~Flynn to the Hipparcos data. For the first time, we apply this method while accounting for parallax and proper motion errors of all individual stars. We do so in a framework of a Bayesian hierarchical model.

In \Sect{sect:meth} we present the general principles of the method, our modelling of the selection function, and how we fit our model to the data, using the information about single stars rather than binned data. In \Sect{sect:res} we present the results. In \Sect{sect:concl} we discuss and conclude.

\section{Method}\label{sect:meth}

In this work we want to model properties of the Galaxy, taking into account all stars' observable properties and astrometric uncertainties. We do so in the framework of a Bayesian hierarchical model (BHM). A BHM is a statistical model with several levels. In this case we have two levels: a top level of `population parameters' and a bottom level of `stellar parameters'. The stellar parameters $\bldpsi_i$ (with $i$ running from 1 to $N$, if our sample is composed of $N$ stars) are a star's true properties (for example its true distance from the Sun, its true velocity, etc.). The population parameters $\bldPsi$ are numbers that characterize the distribution of stars as a whole. Using Bayes theorem, the full posterior on the population parameters $\bldPsi$ and the stellar parameters $\bldpsi = \{\bldpsi_1,\bldpsi_2,...,\bldpsi_N\}$, given the data $\bd = \{\bd_1,\bd_2,...,\bd_N\}$ of all $N$ stars, is 
\begin{equation}\label{eq:Bayes_single}
	\text{Pr}(\bldPsi,\bldpsi | \bd,S) \propto \text{Pr}(\bldPsi) \prod_{i=1}^N
    \frac{S(\bd_i)\text{Pr}(\bd_i | \bldpsi_{i})\text{Pr}(\bldpsi_{i} | \bldPsi)}{\bar{N}(\bldPsi,S)},
\end{equation}
where $\text{Pr}(\bldPsi)$ is a prior on the population parameters, $S(\bd_i)$ is the selection function modelled as acting solely on data, $\text{Pr}(\bd_i | \bldpsi_{i})$ is the likelihood of the data for the $i$-th star (i.e. the pdf of the data $\bd_i$ given the stellar parameters $\bldpsi_i$, sometimes also called the `error model'), $\text{Pr}(\bldpsi_{i} | \bldPsi)$ is the pdf of a star's stellar parameters $\bldpsi_i$ given the population parameters $\bldPsi$, and $\bar{N}(\bldPsi,S)$ a normalization factor (discussed at length in \Sect{sec:norm}). Notice that each star has its own set of stellar parameters $\bldpsi_i$ of some dimension $N_\rmn{sp}$, such that the full space of stellar parameters $\bldpsi$ has dimension $N\times N_\rmn{sp}$.

The stellar parameters themselves are not of much interest to us (they are `nuisance parameters'). Rather, our interest lie in the population parameters $\bldPsi$, one of which will be the total dynamical density in the solar neighbourhood. For these reasons, we want to marginalize over $\bldpsi$ to get a posterior over the population parameters only, i.e.
\begin{equation}\label{eq:Bayes}
	\text{Pr}(\bldPsi | \bd,S) \propto\text{Pr}(\bldPsi)
    \prod_{i=1}^N
    \frac{S(\bd_i)\int \de\bldpsi_{i}\text{Pr}(\bd_i | \bpsi_{i})\text{Pr}(\bpsi_{i} | \bldPsi)}{\bar{N}(\bldPsi,S)}.
\end{equation}
In the rest of this Section, we will analyze each term in the r.h.s. of \Eq{eq:Bayes}.

\subsection{Population model}\label{sect:pop_model}
We express the `population model' $\text{Pr}(\bldpsi_i|\bldPsi)$ as \citep{McMillanBinney2013}
\begin{equation}\label{eq:pop_model}
	\text{Pr}(\bldpsi_i|\bldPsi) \de \bldpsi_i =
    F(M_V) f(\bx,\bv)~\de^3\bx~\de^3\bv~\de M_V,
\end{equation}
so that the position $\bx$, velocity $\bv$, and absolute magnitude $M_V$ are parameterized by a star's stellar parameters $\bldpsi_i$, and $\bldPsi$ represent underlying parameters of the distribution function $f$ and the luminosity function $F$.

\subsubsection{The distribution function $f$}

Let the position $\bx$ in the Galaxy be described by the Cartesian coordinates $(x,y,z)$, with $z$ the height of a star from the Galactic plane positive towards the North Galactic Pole, and $(x,y)$ parallel to the Galactic plane. Let also the velocity $\bv$ be described by $(u,v,w)\equiv(\dot{x},\dot{y},\dot{z})$. 

We assume that the Galaxy is axisymmetric (in \Sect{sect:concl} we will discuss this assumption) and that $\Phi(x,y,z)$ is the Galactic potential. For stars on almost circular orbits we can define $\Phi_z(z)\equiv\Phi(x,y,z)-\Phi(x,y,0)$, and show that the vertical energy $E_z$, defined as 
\begin{equation}
	E_z\equiv\frac{w^2}{2}+\Phi_z(z),
\end{equation}
is an approximate integral of motion \citep{BT2008}. This means that the vertical and the horizontal dynamics of these stars are decoupled, which allows us to factorize the distribution function $f(\bx,\bv)$ as
\begin{equation}
f(\bx,\bv)=f_\rmn{h}(x,y,u,v)\tilde{f}(z,w).
\end{equation}
The stellar number density $\nu(\bx)$ is given by
\begin{equation}
\nu(\bx)=\int\de^3\bv~f(\bx,\bv).
\end{equation}
Defining
\begin{equation}\label{eq:nu}
	\tilde{\nu}(z)=\int_{-\infty}^\infty \de w ~\tilde{f}(z,w),
\end{equation}
and imposing
\begin{equation}\label{eq:norm}
\int_{-\infty}^\infty\tilde{f}(0,w)~\de w=1,
\end{equation}
the stellar number density can be written as
\begin{equation}
\nu(x,y,z)=\nu(x,y,0)\tilde{\nu}(z).
\end{equation}
Because of the Jeans theorem, a population of low eccentricity disc stars with distribution function $\tilde{f}(z,w)=\tilde{f}(E_z(z,w))$ is in statistical equilibrium in the $(z,w)$ space.
Since $E_z=w^2/2$ on the Galactic plane, it is easy to show that for such a distribution function
\begin{equation}\label{eq:vzrelation}
	\tilde{f}(z,w)=\tilde{f}\left(0,\sqrt{w^2+2\Phi_z(z)}\right).
\end{equation}
In other words, the value of $\tilde{f}$ at a certain point $(z,w)$ is given by the value of $\tilde{f}$ on the Galactic plane ($z=0$) for a star with vertical velocity $w'\equiv\sqrt{w^2+2\Phi_z(z)}$. In this way we can rewrite \Eq{eq:nu} as
\begin{equation}\label{eq:nuv2}
	\tilde{\nu}(z)=2\int_{\sqrt{2\Phi_z}}^\infty \de w' ~\frac{\tilde{f}\left(0,w'\right)w'}{\sqrt{{w'}^2-2\Phi_z(z)}}.
\end{equation}
One can select a stellar population supposedly phase-mixed with the potential, fit a function $\tilde{\nu}(z)$ to its vertical distribution and $\tilde{f}(0,w)$ to a subsample of stars close to the Galactic plane. The two functions together allow to derive $\Phi_z(z)$ through \Eq{eq:nuv2}. Using $\Phi_z(z)$ and the approximated Poisson equation (valid close to the Galactic plane)
\begin{equation}\label{eq:Poisson}
	\frac{\partial^2\Phi_z}{\partial z^2}\approx 4\upi G \rho,
\end{equation}
one can derive the total density distribution $\rho(z)$.

As a model for the distribution function on the Galactic plane $\tilde{f}(0,w)$, we use
\begin{equation}
	\tilde{f}(0,w) = \sum_{j=1}^3 C_j(\alpha,\beta)\frac{\exp\left(-\frac{w^2}{2\sigma_j^2}\right)}{\sqrt{2\pi\sigma_j^2}},
\end{equation}
where $\alpha$ and $\beta$ are real numbers in the range $[0,1]$, and $C_1(\alpha,\beta)\equiv\alpha$, $C_2(\alpha,\beta)\equiv(1-\alpha)\beta$, and $C_3(\alpha,\beta)\equiv(1-\alpha)(1-\beta)$. The function $\tilde{f}$ defined in this way respects the normalization condition \Eq{eq:norm}. The distribution function $\tilde{f}(0,w)$ is assumed to be  detailed enough to describe the vertical velocity distribution of stars in the solar neighbourhood \citep[in][they use the sum of two Gaussian distributons]{Creze1998}. Following equation \eqref{eq:vzrelation}, the velocity distribution at height $z$ from the plane is
\begin{equation}\label{eq:fofvz}
	\tilde{f}(z,w) = \sum_{j=1}^3 C_j(\alpha,\beta)\frac{\exp\left(-\frac{w^2+2\Phi_z(z)}{2\sigma_j^2}\right)}{\sqrt{2\pi\sigma_j^2}},
\end{equation}
Integrating this function we obtain the stellar number density
\begin{equation}\label{eq:nuofz}
 	\tilde{\nu}(z)=\sum_{j=1}^3 C_j(\alpha,\beta) \exp\left(-\frac{\Phi_z(z)}{\sigma_j^2}\right).
\end{equation}

\changes{
We model $\Phi_z(z)$ near the Galactic plane as
}

\begin{equation}\label{eq:phiofz}
	\Phi_z(z)=\begin{cases}
    	4\upi G\left(\frac{\rho_0}{2}z^2 + \frac{k}{24}z^4\right), & \text{if $z^2 \leq -\frac{2\rho_0}{k}$},\\
    	4\upi G\left(\frac{2\sqrt{2}}{3}\frac{\rho_0^{3/2}}{\sqrt{-k}}|z| + \frac{\rho_0^2}{2k} \right), & \text{if $z^2 > -\frac{2\rho_0}{k}$}.
 	\end{cases}
\end{equation}
\changes{
with free parameters $\rho_0$ and $k$. The choice of this form is physically motivated by expanding $\rho(z)$ near the Galactic plane up to the 2nd order, and using \Eq{eq:Poisson}. This leads to
}
\begin{equation}\label{eq:rhoofz}
	\rho(z)\approx \rho_0 + \frac{k}{2} z^2+...,
\end{equation}
\changes{
where $\rho_0$ is the total dynamical density in the Galactic plane and $k$ its second derivative. The curvature $k$ is negative, such that the density decreases with distance from the plane. When $z^2 = -2\rho_0/k$, the polynomial expansion of \Eq{eq:rhoofz} reaches a zero density value. In order to avoid unphysical negative densities, the form of the potential at higher absolute values for $z$ increases linearly, corresponding to a constant zero valued density. The form of \Eq{eq:phiofz} satisfies continuity of the potential and its first and second order derivatives for all values of $z$. This issue of strong curvature is largely avoided as our population parameter prior excludes a too steep decrease to the total density with distance from the Galactic plane (see \Sect{sec:prior}).
}

In summary, the stellar number density and the velocity distribution are modelled by 5+2 parameters. These constitute the first 7 of the 9 population parameters $\bldPsi$, listed in the first row of Table~\ref{tab:parameters}. The remaining two population parameters, $z_\odot$ and $W_\odot$, describe the position and velocity of the Sun with respect to the Galactic plane. They are discussed further in \Sect{sec:likelihood_data}.

\subsubsection{The luminosity function $F(M_V)$}

\begin{figure}
	\includegraphics[width=\columnwidth]{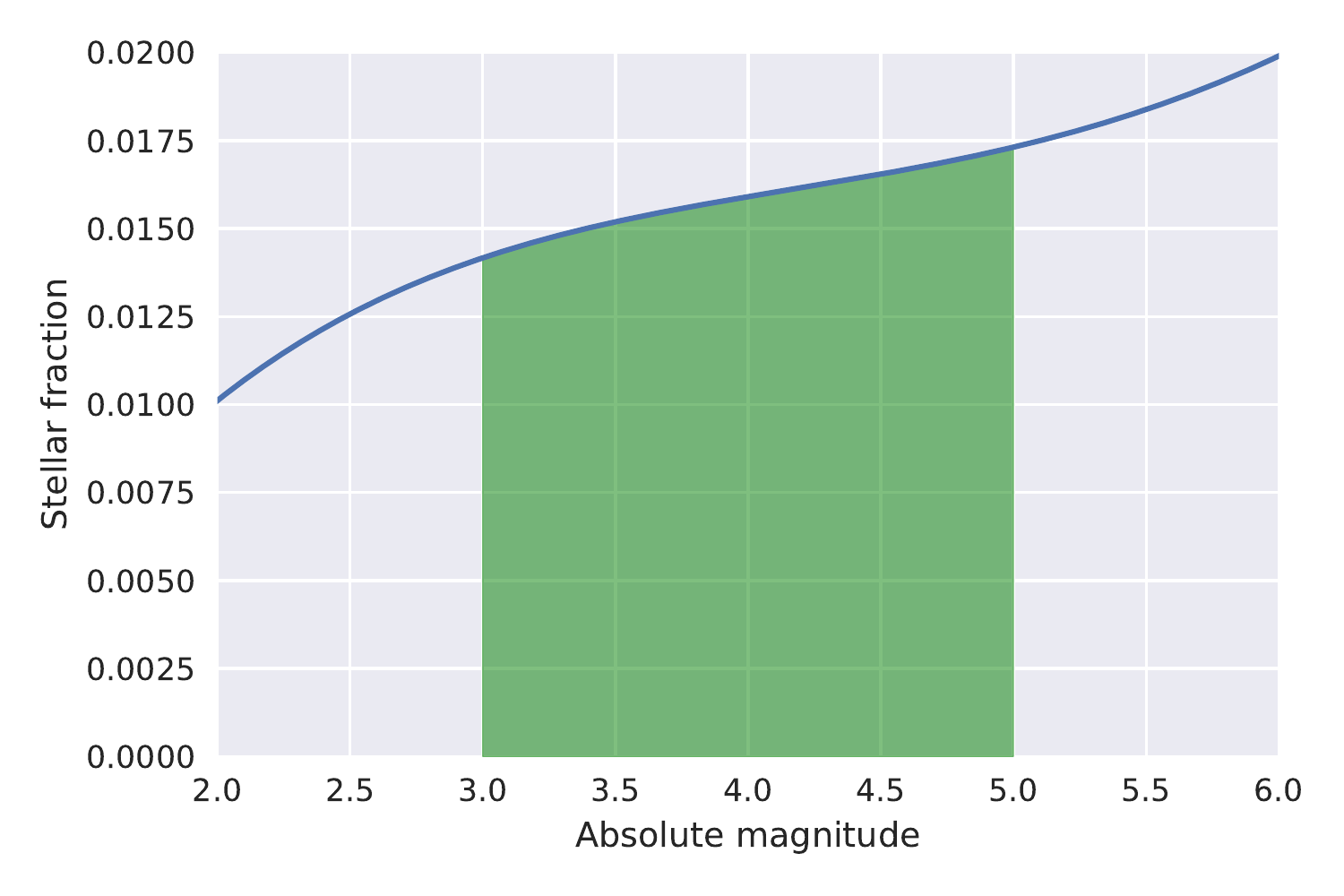}
    \caption{Luminosity function $F(M_V)$, describing the number density of stars in the solar neighborhood, as a function of absolute magnitude. The colored region corresponds to a cut in observed absolute magnitude of sample S1 (see Table~\ref{tab:samples}).}
    \label{fig:P(V)}
\end{figure}

The luminosity function, i.e. the absolute magnitude distribution for stars in the solar neighborhood, is assumed to follow a fifth order polynomial as described in \citep{BinneyMerrifield,McMillanBinney2013}. The distribution is visible in \Fig{fig:P(V)} and, for the region of interest in $M_V$ (see \Sect{sec:samplecuts}), it does not vary dramatically, so for most stars the influence of this factor in the posterior is negligible. However, the function $F(M_V)$ is important for calculating the normalization $\bar{N}$, as we shall see in \Sect{sec:selectionfunction}.

\subsection{Likelihood of the data given the stellar parameters}\label{sec:likelihood_data}

The quantity $\text{Pr}(\bd_i | \bldpsi_{i})$ represents the likelihood of the data $\bd_i$ given a star's stellar parameters $\bldpsi_i$.

A star's data $\bd_i$ consists of astrometric measurements, such as a star's parallax $\varpi$, proper motion in the latitudinal direction $\mu_b$, Galactic longitude $l$ and latitude $b$, and the apparent magnitude $m_V$. We list these data in the second row of Table~\ref{tab:parameters}. The measurements on $l$, $b$ and $m_V$ have negligible uncertainties and can be taken at face value. The observed parallax $\varpi$ and the observed proper motion $\mu_b$, on the other hand, have significant uncertainties that are encoded in the covariance matrix $\Sigma_{\varpi\mu_b}$. These quantities are different from the \emph{true} (or \emph{intrinsic}) parallax and proper motion of a star, which are quantities that would be found if there were no measurement errors. We indicate the true observables with a star symbol, for example $\varpi^{*}$ for the true parallax. The true observables should always be interpreted as functions of the stellar parameters $\bldpsi_i$. The true observables are related to the true stellar position and velocity and to the true absolute magnitude by
\begin{equation}\label{eq:parallax}
	\frac{\varpi^{*}}{\tarcsec}=\sin b \frac{\pc}{z-z_\odot},
\end{equation}
\begin{equation}\label{eq:mub}
	k_\mu \frac{\mu_b^{*}}{\varpi^{*}} = (w - W_\odot)\cos b - [(u-U_\odot) \cos l + (v-V_\odot) \sin l]\sin b,
\end{equation}
\begin{equation}\label{eq:muV}
m_V=M_V+ 5\left[\log_{10}\left(\frac{\tarcsec}{\varpi^{*}}\right)-1\right].
\end{equation}
where $k_\mu=1~\rmn{AU}=4.74057~\kmsec\times\rmn{yr}$, $z_\odot$ is the shift due to the height of the Sun from the Galactic plane, and $W_\odot$ the Sun's vertical velocity; $u$ and $v$ are defined as the star's horizontal Cartesian velocities in the Local Standard of Rest \citep[see][]{BT2008}, with $u$ pointing towards the Galactic centre, $v$ in the sense of the Galactic rotation, and $U_\odot$ and $V_\odot$ the correspondent solar motions. The derivation of $U_\odot$ and $V_\odot$ is a complex problem, outside the scope of this paper, and here we simply use the values derived by \cite{Schonrich2010}, i.e. $U_\odot=11.1~\kmsec$ and $V_\odot=12.24~\kmsec$. However, the Sun's height with respect to the Galactic plane $z_\odot$ and vertical velocity $W_\odot$ are allowed to vary; they are both population parameters, as seen in	 Table~\ref{tab:parameters}.

Notice that the second term of the r.h.s. of \Eq{eq:mub} is very small for stars with low Galactic latitude. Since we do not have access to the $u$ and $v$ components of the velocity (which require the line of sight velocity to be estimated), for low latitude stars ($|b|<20\degr$) we approximate \Eq{eq:mub} as
\begin{equation}\label{eq:mubapprox}
	k_\mu\frac{\mu_b^{*}}{\varpi^{*}} \approx (w - W_\odot)\cos b + [U_\odot \cos l + V_\odot \sin l]\sin b,
\end{equation}
and treat the components due to the horizontal velocities $u$ and $v$ as an extra uncertainty in $\mu_b$ (see below).
\begin{table}
	\centering
	\caption{The 9 population parameters of our model and the data of the respective stars.}
	\label{tab:parameters}
    \begin{tabular}{l}
		\hline
		Population parameters, $\bldPsi$ \\
		\hline
		$\rho_0$,\quad $k$,\quad $\alpha$,\quad $\beta$,\quad $\sigma_1$,\quad $\sigma_2$,\quad $\sigma_3$,\quad $z_\odot$,\quad $W_\odot$ \\
        \hline
        Data, $\bd_{i=1,...,n}$ \\
        \hline
        $\varpi$,\quad $\mu_b$,\quad $\Sigma_{\varpi\mu_b}$,\quad $l$,\quad $b$,\quad $m_V$ \\
		\hline
	\end{tabular}
\end{table}

\subsubsection{Stars with $|b|\leq 20\degr$}

The errors on a star's parallax ($\sigma_\varpi$) and proper motion ($\sigma_{\mu_b}$) are strongly correlated in TGAS. For this reason the catalogue provides also the covariance between the two, $\text{Cov}(\varpi,\mu_b)$. The covariance matrix is
\begin{equation}\label{eq:covariance}
	\Sigma_{\varpi\mu_b} =
    \begin{pmatrix}
    \sigma_\varpi^2	&
    \text{Cov}(\varpi,\mu_b)	\\
    \text{Cov}(\varpi,\mu_b)	&
    \sigma_{\mu_b}^2
\end{pmatrix}
\end{equation}
and unique for each star. We model $\text{Pr}(\bd_i|\bldpsi_i)$ as
\begin{equation}
	\text{Pr}(\bd_i|\bldpsi_i) =\mathcal{N}\left(
\begin{pmatrix}
\varpi^{*}(\bldpsi_i) \\
\mu_b^{*}(\bldpsi_i)
\end{pmatrix},
\begin{pmatrix}
\varpi \\
\mu_b
\end{pmatrix},
\tilde{\Sigma}_{\varpi\mu_b}\right),
\end{equation}
where $\mathcal{N}(\bmath{u},\bar{\bmath{u}},\Sigma)$ is the multivariate distribution of $\bmath{u}$, with mean $\bar{\bmath{u}}$, and covariance matrix $\Sigma$. Here we use an {\it effective} covariance matrix
\begin{equation}\label{eq:covariancetilde}
	\tilde{\Sigma}_{\varpi\mu_b} =
    \begin{pmatrix}
    \sigma_\varpi^2 + \sigma_{\varpi,\rmn{sys.}}^2	&
    \text{Cov}(\varpi,\mu_b)	\\
    \text{Cov}(\varpi,\mu_b)	&
    \sigma_{\mu_b}^2 + \left(\frac{\varpi\sin b\, \sigma_x}{k_\mu}\right)^2
\end{pmatrix}.
\end{equation}
The tilde in $\tilde{\Sigma}_{\varpi\mu_b}$ signifies that the covariance matrix is modified, in that it contains two additional sources of uncertainty: $\sigma_{\varpi,\rmn{sys.}}$ and $\sigma_x$. The term $\sigma_{\varpi,\rmn{sys.}} = 0.3~\mas$ is the systematic component that should be added to the parallax uncertainty according to \cite{GaiaDR1}. The additional component including $\sigma_x$ is our way to model the effects of a star's horizontal $u$ and $v$ motions -- the second and third term in the r.h.s. of \Eq{eq:mub}. We choose $\sigma_x=40~\kmsec$ \citep[e.g.][Table 10.2]{BinneyMerrifield}. This crude modelling of the horizontal kinematics is sufficient for our aim to describe the vertical kinematics of stars at low latitude, which in our case we define as the stars with $|b| \leq 20\degr$. \changes{This also accounts for possible uncertainty in the Sun's velocity parallel to the plane ($U_\odot$ and $V_\odot$); the uncertainties on these quantities are small \citep[$<1~\kmsec$ in][]{Schonrich2010} and negligible compared to the velocity dispersion $\sigma_x$.}

Our limiting angle of $20^\circ$ is larger than what have been used in previous studies of the same method. Unlike these studies, we account for horizontal velocity effects to the proper motion and also how the velocity distribution changes with $z$, which is why we can be less restrictive in terms of this limiting angle.

\subsubsection{Stars with $|b|>20\degr$}

For stars at high latitudes, the terms due to the horizontal velocities in \Eq{eq:mub} become important, and it becomes increasingly difficult to infer the vertical velocity. We therefore decide to ignore the kinematic information for stars with Galactic latitude $|b|>20\degr$. The likelihood $\text{Pr}(\bd_i|\bldpsi_i)$ simplifies to 
\begin{equation}\label{eq:Gaussian}
\text{Pr}(\bd_i|\bldpsi_i)=G\left(\varpi^{*}(\bldpsi_i),\varpi,\sqrt{\sigma_\varpi^2 + \sigma_{\varpi,\rmn{sys.}}^2}\right),
\end{equation}
where $G(u,\bar{u},\sigma)$ is the Gaussian distribution of $u$, with mean $\bar{u}$, and standard deviation $\sigma$. 

Since the integrals in \Eq{eq:Bayes} do not depend anymore on the kinematic information, for these stars we can marginalize over the velocity $\bv$ and $\text{Pr}(\bpsi_i|\bldPsi)$ becomes
\begin{equation}
\text{Pr}(\bldpsi_i|\bldPsi) \de \bldpsi_i =
    F(M_V) \nu(\bx)~\de^3\bx~\de M_V.
\end{equation}

\subsection{Selection function}
\label{sec:selectionfunction}

The TGAS catalogue is incomplete in many respects. In Gaia DR1, different parts of the sky have been scanned a different number of times. An individual star needs to be observed and identified at least 5 times to be included in DR1; due to a complicated scanning pattern, some parts of the sky have significantly worse coverage than others. Guided by the principle that it is better to have a large statistical uncertainty rather than an unknown bias, we mask large parts of the sky where selection effects are severe and difficult to model. In the non-masked region we calculate the completeness of the TGAS catalogue. Finally, we make sample cuts in absolute magnitude and distance.

We model the selection function as a function solely on data. Because of this, the selection factor $S(\bd_i)$ of a star is constant, as the data is fixed. Importantly, the selection function is essential for calculating the normalization $\bar{N}(\bldPsi,S)$, as we shall see in Section \ref{sec:norm}.

\subsubsection{Masking}
\label{sec:masking}

We bin the sky into 3072 pixels and apply the following selection criteria.

\begin{figure}
	\includegraphics[width=\columnwidth]{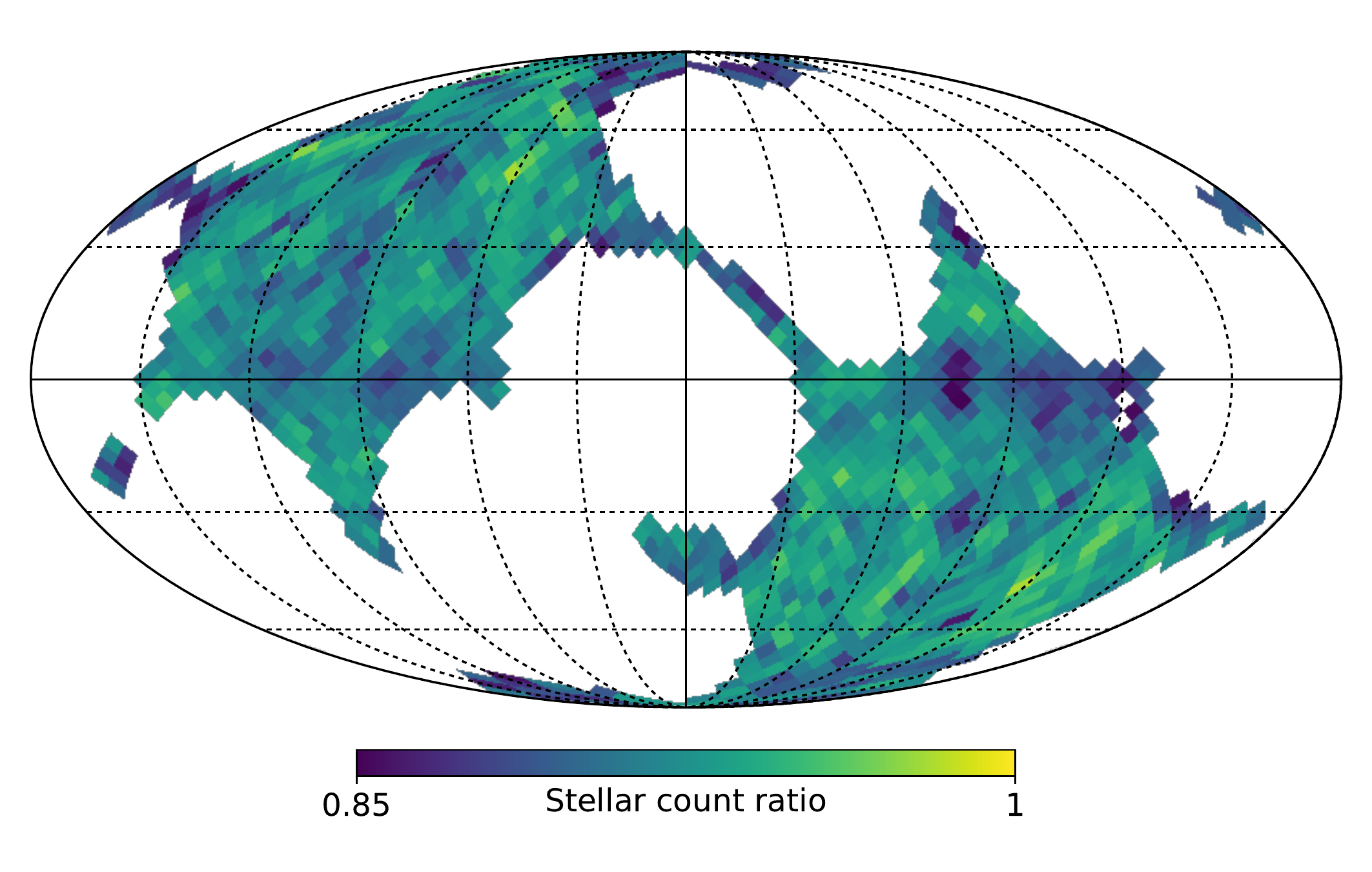}
    \caption{Mollweide projection showing the ratio of stellar counts between TGAS and Tycho-2 for stars with apparent magnitude in range $m_V\in [7,11]$. Masked pixels, outside the region $I_\Omega$, are white. The centre point of the figure has coordinates $(l,b)=(0,0)$, with $l$ increasing to the left and $b$ increasing upwards in the figure.}
    \label{fig:completeness_map}
\end{figure}

\begin{itemize}
	\item The Tycho-2 catalogue is said to be 99 per cent complete up to magnitude $m_V=11$. For each pixel we calculate the stellar count ratio of stars with apparent magnitude in range $m_V\in [7,11]$, between the TGAS and Tycho-2 catalogues. We mask any pixel for which this ratio is less than 85 per cent. We also apply the same criteria to a larger area around the pixel, which includes all adjacent pixels. If this 9 pixel region has a stellar count ratio that is less than 85 per cent we mask the center pixel of that region.
	\item Some regions have been scanned only a few times by the Gaia telescope, producing incompleteness effects and biases that are very difficult to account for. These regions can be masked by considering the number of independent observations of stars in a pixel. We do not want pixels with stars that have a significant lower tail of very few independent observations. If the number of independent observations of stars in a pixel has a 20th percentile strictly lower than 9.5 or a 10th percentile strictly lower than 7.5, we mask that pixel and all its immediate neighbours.
    \item In order to remove small islands of unmasked pixels, we mask any pixel that has 6 or more already masked neighboring pixels.
\end{itemize}
These criteria masks 58 per cent of the sky, as it is visible in \Fig{fig:completeness_map}. This constitutes a first component to the selection function, $S\propto\theta(\{l,b\}\in I_\Omega)$, where $\theta$ is a step function and $I_\Omega$ is the non-masked region of the sky.

\subsubsection{Completeness}
\label{sec:completeness}

\begin{figure}	\includegraphics[width=\columnwidth]{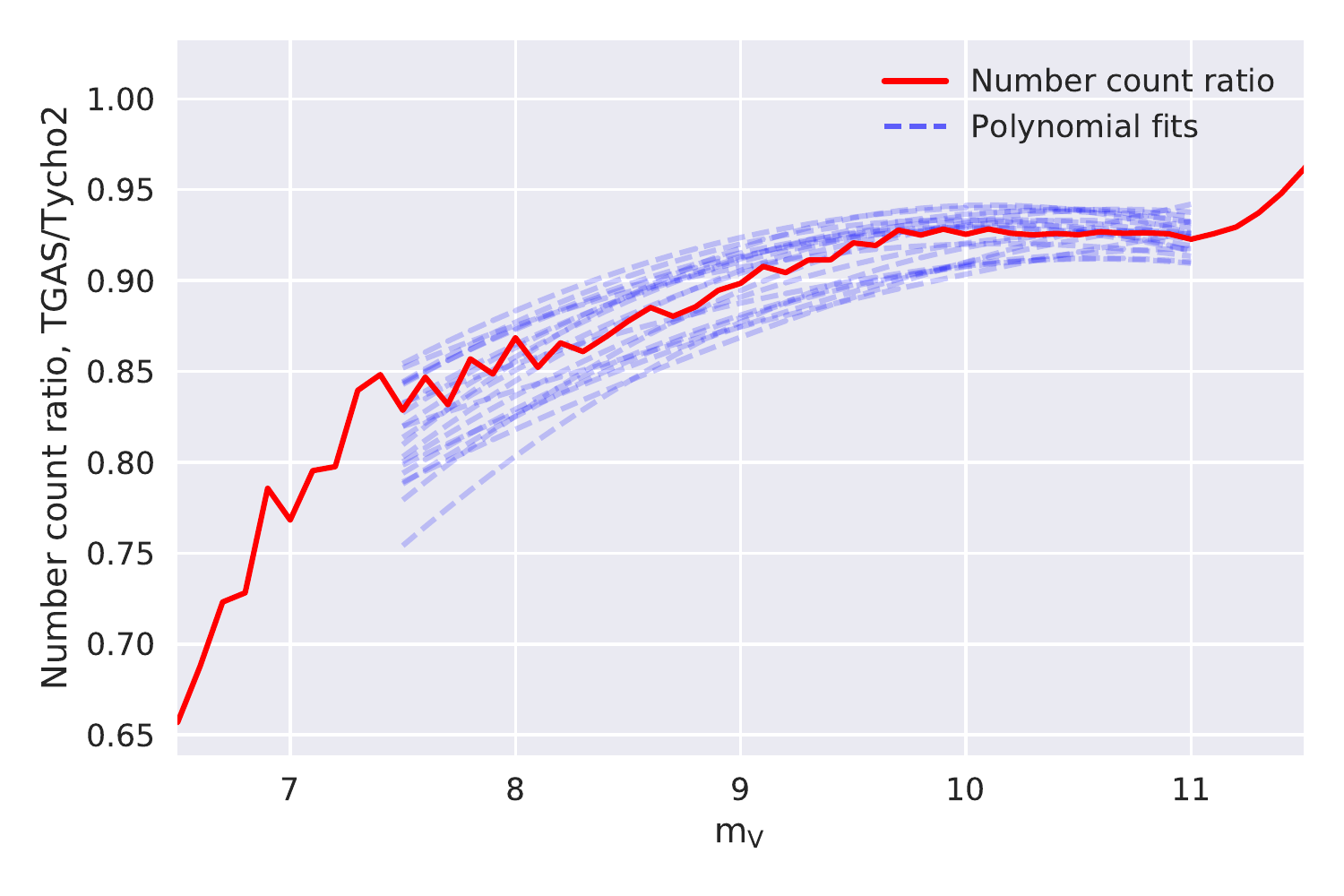}
    \caption{Completeness as a function of apparent magnitude $m_V$. The solid red line represents the ratio between the TGAS and Tycho-2 catalogue, over the full non-masked region of the sky. The transparent dashed blue lines are second order polynomial fits to the same quantity, for 20 different bins in Galactic latitude $b$.}
    \label{fig:selection_V}
\end{figure}

In \Fig{fig:selection_V}, we show the relative completeness between the TGAS and Tycho-2 catalogues over the unmasked region, as a function of the apparent magnitude $m_V$. As is apparent from this figure, Gaia DR1 is missing a lot of brighter objects, especially for $m_V<7.5$. For objects dimmer than $m_V=11$, Tycho-2 is known to have increasingly poorer completeness. We approximate the Tycho-2 catalogue to be complete for magnitudes $m_V<11$ and constrain ourselves to use objects in the region $m_V\in [7.5,11.0]$. In order to account for a possible dependence on $b$ in the selection function, we bin the unmasked sky in 20 subregions of $b$, with an approximately equal number of stars in each bin. For each of these bins in $b$, we fit a second order polynomial to the ratio of stars between TGAS and Tycho2, as it is visible in \Fig{fig:selection_V}. We do not need to account for any dependence on the Galactic longitude $l$, as it cannot affect $\bar{N}(\bldPsi,S)$ or our end result (see Section~\ref{sec:norm}). Thus we get a completeness function for the non-masked part of the sky, $S\propto C(m_V,b)$. This can equally well be interpreted as a function of data or of intrinsic stellar properties, as the errors on these observables are negligible and disregarded in our model.

\subsubsection{Sample cuts}
\label{sec:samplecuts}

In order to have a stellar sample to work with, we need to make some cuts in observable quantities. As mentioned in \Sect{sec:completeness}, the completeness is sufficient and well described only in a limited range of apparent magnitudes. We also want to limit ourselves in distance, as many of our assumptions are only valid in the solar neighbourhood.

If we were to make cuts in distance and apparent magnitude, we would risk to bias our analysis by looking at different populations of stars: stars in the sample that are close would be dim, while stars far away would be bright. In such a case, dim stars would constrain the velocity distribution, as the data for close stars is better, while bright stars would constrain the stellar number density at high $|z|$. This is problematic as the populations of bright and dim stars might be different. Furthermore, the result would be highly degenerate with the shape of the luminosity function $F(M_V)$. We prefer to use a method and sample that does not heavily rely on a priori knowledge about the stellar population and is not in danger of biases coming from looking at different stellar populations.

Ideally, we would like to construct our sample by making a cut in intrinsic absolute magnitude. Of course, the intrinsic absolute magnitude is not known, so we have to resort to the next best thing: the \emph{observed} absolute magnitude,
\begin{equation}\label{eq:observedabsmagnitude}
	\MVobs = m_V - 5\Big[\log_{10}\Big(\frac{\tarcsec}{\varpi}\Big)-1\Big].
\end{equation}
We also need to constrain ourselves in distance by making a cut in observed parallax. By having both upper and lower bounds to the observed parallax, we can at the same time ensure that all stars have apparent magnitudes in our preferred region $m_V\in [7.5,11.0]$. With these cuts to the data, the minimum and maximum values of the apparent magnitude are given by
\begin{equation}
\begin{split}
	\min(m_V) = \min(\MVobs) + 5\Big[\log_{10}\Big(\frac{\tarcsec}{\max(\varpi)}\Big)-1\Big], \\
    \max (m_V) = \max(\MVobs) + 5\Big[\log_{10}\Big(\frac{\tarcsec}{\min(\varpi)}\Big)-1\Big].
\end{split}
\end{equation}
All samples used in this articles are listed in Table~\ref{tab:samples}, where $I_{M_V}$ is the interval of observed absolute magnitude and $I_\varpi$ is the interval of observed parallax by which we construct our samples. Also visible in the table is the distance interval corresponding to the parallax values. In should be noted that the true distance of the object can actually lie outside this interval, due to errors in the parallax measurement.
\begin{table}
	\centering
	\caption{List of the stellar samples used in this work. The columns are: sample name, the interval of observed absolute magnitude, the interval of observed parallax, the distance corresponding to these parallax values, the number of stars in a sample.}
    \label{tab:samples}
	\begin{tabular}{r || l l l l}
		\hline
		Sample & $I_{M_V}$ & $I_\varpi~(\mas$) &  Dist. $(\rmn{pc})$ &  \# of stars \\
		\hline
        S1 & 3--5 & 6.25--12.5 & 80--160 &  19,350 \\
		\hline
        S2 & 4--5.5 & 8--20 & 50--125 & 9,737 \\
		\hline
        S3 & 4--4.5 & 5--20 & 50--200 & 11,781 \\
		\hline
        S4 & 3--4 & 4--12.5 & 80--250 & 28,614 \\
		\hline
	\end{tabular}
\end{table}

These cuts gives a third and final component to our selection function, $S\propto\theta(\MVobs\in I_{M_V})\theta(\varpi\in I_\varpi)$,
which require that the observed absolute magnitude and parallax are in their allowed regions.

\subsubsection{Normalization}
\label{sec:norm}

The number of expected stars in a sample changes when varying the population parameters, making it necessary to compute the normalization factor $\bar{N}(\bldPsi,S)$ (seen in \Eq{eq:Bayes_single} and \Eq{eq:Bayes}).

The expected number of stars in a sample is an integral over the distribution of intrinsic stellar properties, convolved with errors, and the probability of being selected. It is written

\begin{equation}\label{eq:Nbar_theory}
	\bar{N}(\bldPsi,S) =
    \int S(\diprime) \text{Pr}(\diprime | \bpsi_i') \text{Pr}(\bpsi_i'|\bldPsi) \de\bpsi_i' \de\diprime.
\end{equation}
Note that this normalization has nothing to do with the actual observations of our sample, but is an integral over \emph{generated} data, given the population model and the observational errors of the survey. This is signified by the prime in $\diprime$ and $\bpsi_i'$, meaning that we are integrating over the possible stellar parameters and data of a hypothetical star.

The selection function $S(\diprime)$ is a function solely on data (although the data is equivalent to the true observables for angular position on the sky and apparent magnitude). It is in three parts: one part coming from masking a large fraction of the sky, one part coming from the TGAS survey not being complete in the non-masked region, and one part coming from our sample cuts, as explained in Sections~\ref{sec:masking}, \ref{sec:completeness} and \ref{sec:samplecuts}. The full selection function is
\begin{equation}
	S(\diprime) = \theta(\{l,b\}\in I_\Omega)C(m_V,b)\theta(\MVobs \in I_{M_V})\theta(\varpi \in I_\varpi),
\end{equation}
where $I_\Omega$ is the non-masked region of the sky, $C(m_V,b)$ is the completeness in the non-masked region, and $I_{\varpi}$ and $I_{M_V}$ are the regions of observed parallax and absolute magnitude by which we construct our sample.

\begin{figure}
	\includegraphics[width=\columnwidth]{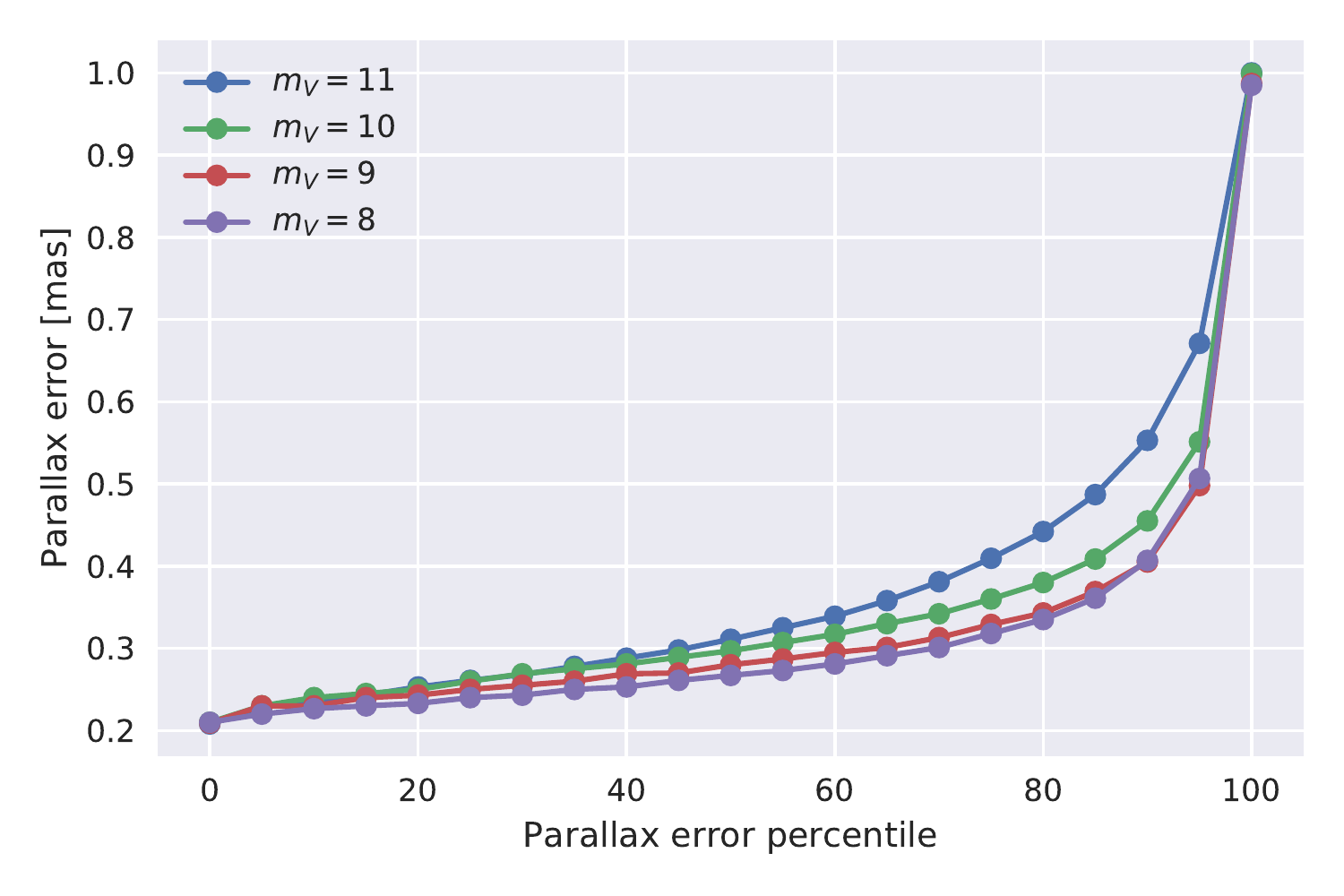}
    \caption{Parallax uncertainty percentiles for stars in the non-masked region of the sky, with apparent magnitude $m_V = 8,9,10,11$. Evidently, dimmer objects have slightly larger parallax errors.}
    \label{fig:parallax_error_percentiles}
\end{figure}

The factor $\text{Pr}(\diprime | \bpsi_i')$ is the probability that a hypothetical star with stellar parameters $\bpsi_i'$ has observed data $\diprime$. The true parallax is the only observable quantity with significant uncertainty that has an effect on whether or not the star is selected. All other observables are either delta functions or can be integrated out to one, as in the case of the proper motion. The likelihood of the data simplifies to
\begin{equation}\label{eq:likelihood_int}
	\text{Pr}(\diprime | \bpsi_i') =
    \int \de\sigma_\varpi g(\sigma_\varpi,m_V)
    G\left(\varpi^{*}(\bldpsi_i),\varpi,\sigma_\varpi\right),
\end{equation}
where $\varpi^{*}(\bpsi_i')$ is the true parallax as given by the star's true position, and $g(\sigma_\varpi,m_V)$ is the distribution of parallax uncertainties. We take $g(\sigma_\varpi,m_V)$ to be the distribution of parallax uncertainties for actual stars in the TGAS catalogue that are in the non-masked region of the sky. This distribution is visible in \Fig{fig:parallax_error_percentiles}, represented in terms of percentiles. The uncertainties range from $0.2$ to $1~\mas$ and vary somewhat with $m_V$. No significant dependence on angular position was found, which is why we write the distribution of parallax uncertainties as $g(\sigma_\varpi,m_V)$, also adding in quadrature the systematic parallax uncertainty $\sigma_{\varpi,\rmn{sys.}}=0.3 ~\mas$.

Finally, since the velocity is unimportant for the definition of our stellar samples, the probability of a star's intrinsic position and intrinsic absolute magnitude simplifies to
\begin{equation}
	\text{Pr}(\bpsi_i'|\bldPsi)\de\bpsi_i'  = 
    \nu(\bx) F(M_V) \de^3\bx \de M_V.
\end{equation}

Spelled out in detail, equation \eqref{eq:Nbar_theory} becomes
\begin{equation}\label{eq:numberofstars}
\begin{split}
	\bar{N}(\bldPsi,S) =& 
    \int \de^3\bx\nu(\bx) \theta(\{l,b\}\in I_\Omega) \\
   & \int \de M_V C(m_V,b) F(M_V) \int \de\sigma_\varpi g(\sigma_\varpi,m_V) \\
   & \int \de\varpi G\left(\varpi^{*}(\bldpsi_i),\varpi,\sigma_\varpi\right)\theta\left(\MVobs \in I_{M_V}\right)\theta(\varpi \in I_\varpi),
\end{split}
\end{equation}
now written as a nested integral. It is implicit that the angles $l$ and $b$ are given by the true position $\bx$ and $z_\odot$, that $\varpi^{*}$ is given by the true distance, and that the apparent magnitude $m_V$ is a function of the true absolute magnitude and true distance. In the volume integral, the range of integration includes only the volume which is part of the non-masked sky, $I_\Omega$. In the integral over observed parallax, the range of integration is the interval which satisfies the criteria that $\varpi \in I_{\varpi}$ and $\MVobs\in I_{M_V}$.

\begin{figure}
	\includegraphics[width=\columnwidth]{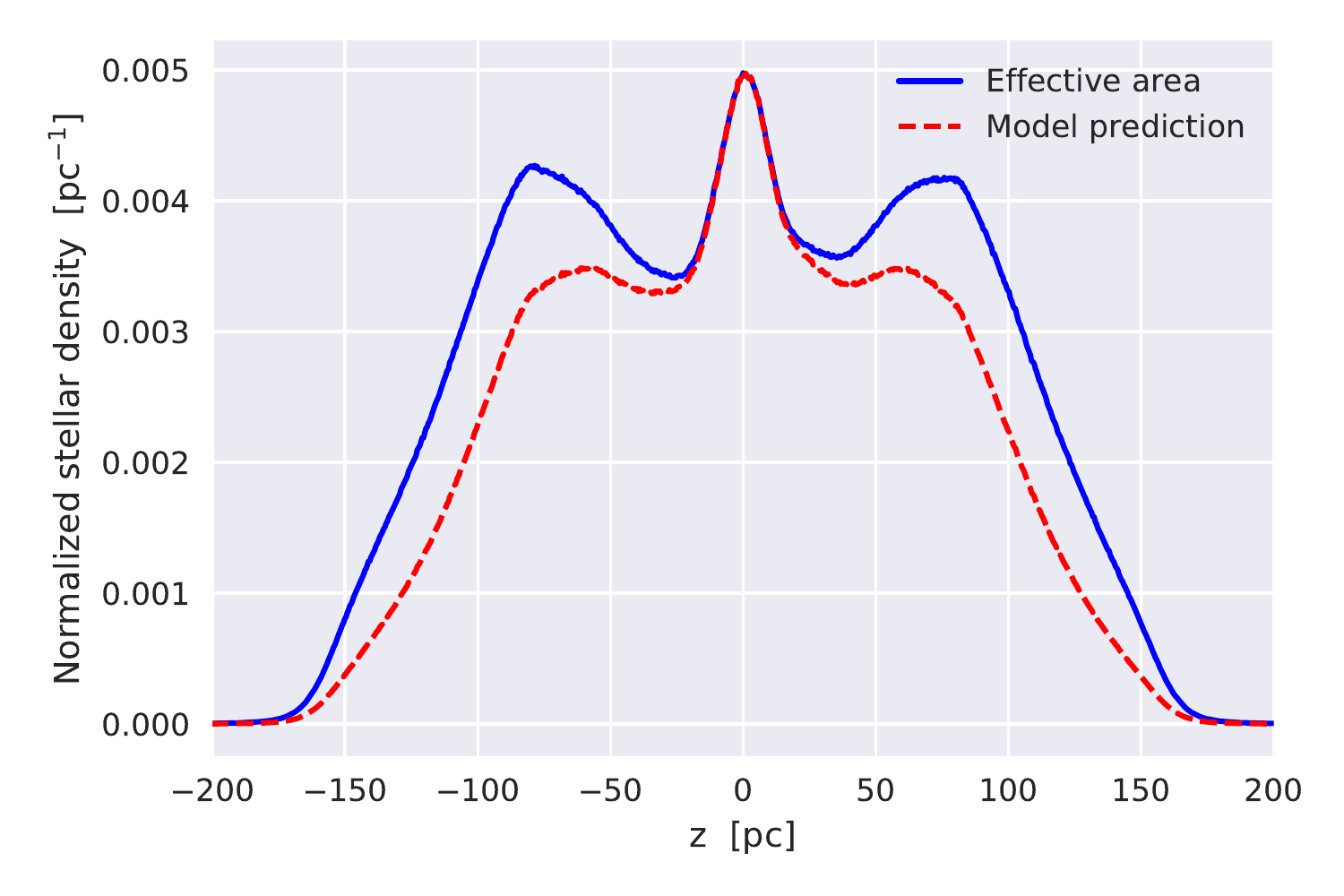}
    \caption{Effective area (solid blue) and model prediction (dashed red) for sample S1. The model prediction of the stellar counts is $\tilde{\nu}(z)A_\text{eff}(z)$, where $\tilde{\nu}(z=0)=1$. The population parameters are $\rho_0=0.1~\Msun\pc^{-3}$, $\alpha=0.5$, $\beta=0.9$, $\sigma_{\{1,2,3\}}=\{6,14,30\}~\kmsec$ and other parameters set to zero. The effective area is normalized such that the integral over $z$ produces unity. Clearly, the normalization factor $\bar{N}(\bldPsi,S)$ becomes smaller with higher density $\rho_0$.}
    \label{fig:effective_area}
\end{figure}

Evaluating the integral of \Eq{eq:numberofstars} is very costly, but can be made much faster by cataloging a vector of effective areas at different heights in the solar frame, $A_\text{eff}(z-z_\odot)$. In this way the integral becomes one-dimensional over $z$, i.e.,
\begin{equation}\label{eq:Nnu}
	\bar{N}(\bldPsi,S) = \nu(x,y,0)\int \tilde{\nu}(z)A_\text{eff}(z-z_\odot)~\de z
\end{equation}
where the effective area $A_\text{eff}(z-z_\odot)$ is given by
\begin{equation}\label{eq:Aeff}
\begin{split}
	A_\rmn{eff}(z-&z_\odot) = \int \de x \de y \theta(\{l,b\}\in I_\Omega) \\
    &\int \de M_V C(m_V,b) F(M_V) \int \de\sigma_\varpi g(\sigma_\varpi,m_V) \\
    &\int \de\varpi G\left(\varpi^{*}(\bldpsi_i),\varpi,\sigma_\varpi\right)\theta\left(\MVobs \in I_{M_V}\right)\theta(\varpi \in I_\varpi).
\end{split}
\end{equation}
In \Eq{eq:Nnu}, we assume that $\nu(x,y,0)$ is approximately constant in the small volumes probed by our stellar samples, becoming an unimportant constant that goes out of the integral and simplifies in \Eq{eq:Bayes}, because it appears both at the numerator and at the denominator.
The effective area of one of the samples used in this work is shown in solid blue in \Fig{fig:effective_area}. This would correspond to the distribution of stars in the sample, if the stellar number density was uniform in space. For comparison, the dashed red lines in the same figure represents $\tilde{\nu}(z)A_\rmn{eff}(z)$, assuming an example population model.

\subsection{Prior}
\label{sec:prior}

The factor $\text{Pr}(\bldPsi)$ is a prior on the population parameters. We require that $\rho_0$ is positive, $k$ negative, that $\sigma_{1,2,3}$ are positive and in ascending order, and that $\alpha$ and $\beta$ are in range $[0,1]$. In addition to these criteria, the prior has the form
\begin{equation}\label{eq:p-p-prior}
	\text{Pr}(\bldPsi) \propto
    \exp\left( -\frac{1}{2}\frac{|k|}{\rho_0}\frac{\kpc^2}{25} \right),
\end{equation}
where $|k|/\rho_0 = 25~\kpc^{-2}$ corresponds to a density that falls off to half its value at $z=200~\pc$ (and zero density at $z \sim 283~\pc$). This prior means that we regard with skepticism densities that fall off too fast with $z$.

\subsection{MCMC sampling}
\label{sec:sampling}

We want to calculate the posterior on the population parameters with the stellar parameters marginalized, as written in \Eq{eq:Bayes}. The conceptually straightforward procedure would be to run a Monte-Carlo Markov Chain (MCMC) over this function, in the parameter space of the 9 population parameters. However, the integrals over the stellar parameters are computationally very expensive. In order to get around this problem we actually compute the posterior in the full space of population parameters and stellar parameters alike, giving a total of $9+N \times N_\rmn{p}$ dimensions ($N \simeq 10^4$ is the number of stars in our samples, $N_p$ is the number of stellar parameters per object). Even though the number of dimensions is hugely increased, the posterior can actually be computed efficiently with a special purpose MCMC sampling method called Metropolis-within-Gibbs \citep{BayesianDataAnalysis}.

Much like the BHM formalism, as expressed in \Eq{eq:Bayes_single}, the Metropolis-within-Gibbs MCMC sampler operates on two levels: it alternates between varying the stellar parameters with the population parameters fixed, and varying the population parameters with the stellar parameters fixed. There is no correlation between the stars in our model; they are assumed to be independently drawn from the population distribution. Therefore the random walk of the stellar parameters is performed for one star at a time. This is very useful, as it is much cheaper to retrieve a statistically independent point from $n$ random walks in $m$ dimensions, rather than one random walk in $n \times m$ dimensions.

Breaking it down, the Metropolis-within-Gibbs MCMC sampler goes through the following steps:

\begin{enumerate}
	\item Take some number of steps in stellar parameter space (population parameters fixed), using a Metropolis random walk. This is done for each star separately.
    \item Fix the stellar parameters to the final parameter values of that random walk.
    \item Take some number of steps in the population parameter space (stellar parameters fixed), using a Metropolis random walk.
    \item Fix the population parameters to the final parameter values of that random walk.
    \item Append the population parameter values to an output chain.
    \item Repeat from (i).
\end{enumerate}
In (i) and (iii), it is important that the random walk takes sufficiently many steps, such that the end points of the respective Metropolis random walks are statistically independent from their starting points. If the step length of the random walk is well calibrated, a statistically independent point is retrieved after a small number of steps.

By using this sampling technique, the stellar parameters are marginalized over in the output chain. It is computationally efficient; even though we are moving through a parameter space of order $10^4$ dimensions, we are only varying 9, 2 or 1 parameters at any one time.

There is a computational simplification coming from the fact that we can drop any constant terms when evaluating posterior values for the Metropolis random walk of steps (i) and (iii). When varying the stellar parameters, the normalization $\bar{N}(\bldPsi,S)$ does not vary, so any selection function effects do not have to be computed. When varying the population parameters, the likelihood of the data can be dropped, as it depends only on the stellar parameters. This is explained in detail below.

\subsubsection{Varying the stellar parameters}
\label{sec:stellarparameters}

Because we are only considering observational errors on the parallax and proper motion, the stellar parameters only need to specify the distance and velocity perpendicular to the plane. We parameterize the intrinsic stellar properties by stellar parameters $s$ and $W$, which are distance and velocity in $z$-direction in the Sun's rest frame. These are related to $z$ and $w$ by 
\begin{equation}\label{eq:z}
	z = s\sin b+z_\odot,
\end{equation}
and
\begin{equation}\label{eq:w}
	w = W+W_\odot.
\end{equation}

Dropping any factors that do not depend on the stellar parameters, the posterior density of \Eq{eq:Bayes_single} is proportional to
\begin{equation}
	\text{Pr}(\bldPsi,\bldpsi | \bd,S) \propto \prod_i \text{Pr}(\bd_i | \bldpsi_i)\text{Pr}(\bldpsi_i | \bldPsi).
\end{equation}
Disregarding the velocity information for stars with $|b|\geq 20^\circ$, the posterior is proportional to
\begin{equation}\label{eq:stellarparameters1}
\begin{split}
	\text{Pr}(\bd_i | \bldpsi_i)\text{Pr}(\bldpsi_i | \bldPsi) \propto G\left(\varpi^*(s),\varpi,\sqrt{\sigma_\varpi^2+\sigma_{\varpi,\text{sys}}^2}\right)\\
   \times ~ s^2 F(M_V)\tilde\nu\left(z(s,z_\odot)\right),
\end{split}
\end{equation}
where the factor $s^2$ comes from the volume element Jacobian and
\begin{equation}
	\frac{\varpi^*}{\rmn{arcsec}}=\frac{\rmn{pc}}{s},
\end{equation}
\begin{equation}
	M_V = m_V - 5\left[\log_{10}\left(\frac{s}{\text{pc}}\right)-1\right].
\end{equation}

For stars with a Galactic latitude $|b|<20^\circ$, for which we also consider the proper motion information, we have
\begin{equation}\label{eq:stellarparameters2}
\begin{split}
	\text{Pr}(\bd_i | \bldpsi_i)\text{Pr}(\bldpsi_i | \bldPsi) \propto\mathcal{N}\left(
\begin{pmatrix}
\varpi^{*}(s) \\
\mu_b^{*}(s,W)
\end{pmatrix},
\begin{pmatrix}
\varpi \\
\mu_b
\end{pmatrix},
\tilde{\Sigma}_{\varpi\mu_b}\right) \\ \times ~ s^2 F(M_V) \tilde{f}\left(z\left(s,z_\odot\right),w\left(W,W_\odot\right)\right). \\
\end{split}
\end{equation}
The distributions $\tilde{\nu}$ and $\tilde{f}$ are given by equations \eqref{eq:fofvz}, \eqref{eq:nuofz} and \eqref{eq:phiofz}.

\subsubsection{Varying the population parameters}
\label{sec:populationparameters}

When varying the population parameters we can drop any terms that depend only on the data and/or the stellar parameters. Doing so, the posterior probability is proportional to
\begin{equation}
	\text{Pr}(\bldPsi,\bldpsi | \bd,S) \propto \text{Pr}(\bldPsi) \prod_i \frac{\text{Pr}(\bldpsi_i | \bldPsi)}{\bar{N}(\bldPsi,S)},
\end{equation}
where, for $|b|\geq 20^\circ$,
\begin{equation}
	\text{Pr}(\bldpsi_i | \bldPsi) \propto \tilde{\nu}\left(z\left(s,z_\odot\right)\right)
\end{equation}
and, $|b| < 20^\circ$,
\begin{equation}
	\text{Pr}(\bldpsi_i | \bldPsi) \propto
    \tilde{f}\left(z\left(s,z_\odot\right),w\left(W,W_\odot\right)\right).
\end{equation}
The final part of the posterior is the normalization $\bar{N}(\bldPsi,S)$, as given by equation \eqref{eq:numberofstars}.

\subsubsection{Burn-in}
\label{sec:burn-in}

The steps of the MCMC random walk are drawn from multivariate Gaussian distribution; we have a 9-dimensional covariance matrix for the population parameters, and a 1- or 2-dimensional covariance matrix (depending whether the star's $|b|$ is larger or smaller than $20^\circ$) for each star. To make the sampler efficient, the step length covariance matrices need to be calibrated such that a statistically independent sample is retrieved after a minimum number of attempted steps. These matrices are determined in a careful and thorough burn-in phase. The burn-in consists of several parts, such that the sampler is allowed to find the peak of the posterior and then calibrate its step lengths in that region.
The burn-in phase consists in the following steps:
\begin{itemize}
\item a burn-in of the stellar parameters (population parameters fixed), in 200 steps and then 500 steps, setting a preliminary step length to the stellar parameters;
\item a Metropolis-within-Gibbs random walk of 10 steps in population parameter space and 10 steps in stellar parameter space, in 100 iterations (such that the sampler finds the peak of the posterior density, without changing any step length covariance matrices);
\item a burn-in of the population parameters (stellar parameters fixed), in 500 steps and then 1000 steps, setting a preliminary step length to the population parameters;
\item a burn-in of the stellar parameters (population parameters fixed) of 1000 steps, setting the final step length covariance matrices of the stellar parameters;
\item a Metropolis-within-Gibbs burn-in of 20 steps in population parameter space and 10 steps in stellar parameter space, in 500 iterations, setting the final step length covariance matrix of the population parameters.
\end{itemize}
This last step sets the covariance matrix of the population parameters, although not as a covariance on the population parameter chain. Rather, the covariance matrix is set as a covariance on the difference in population parameters between iterations. In other words, if $\{x_k\}$ is the chain of population parameters, one point $x_k$ from each iteration of the Metropolis-within-Gibbs burn-in, the covariance matrix on the population parameters is not $\text{Cov}(\{x_k\})$, but instead $\text{Cov}(\{x_{k+1}-x_k\})$. This is because the sampler is a Gibbs sampler in the highest level and the posterior density is not separable in population and stellar parameters; even though the Metropolis sampler draws a statistically independent sample of stellar parameters given fixed population parameters, and vice versa, the population parameter values from two neighboring elements of the output chain are not necessarily statistically independent. Therefore, the efficient step length in the population parameters space is computed from difference in population parameters between neighbouring iterations.

\changes{
In order to support that our sampling algorithm, error handling and treatment of selection effects are viable, we demonstrate our method on mock data. This can be found in Appendix~\ref{app:toydata}.
}

\section{Results}\label{sect:res}

As is detailed in \Sect{sec:sampling}, the posterior density distribution of the population parameters $\bldPsi$, marginalized over the stellar parameters $\bldpsi$, is found by a Metropolis-within-Gibbs Monte-Carlo Markov Chain (MCMC). This is done for four different samples of stars, listed in Table~\ref{tab:samples}. The posterior distributions are calculated by a Metropolis-within-Gibbs MCMC of 20,000 iterations. Each iteration consists of a 20-step Metropolis random walk in stellar parameter space and a 40-step Metropolis random walk in population parameter space.
\begin{table}
	\centering
	\caption{Posterior results of the dynamical matter density and the Sun's position and velocity with respect to the plane, presented as the median plus/minus its 16th and 84th percentiles.}
    \label{tab:results}
	\begin{tabular}{r || l l l}
		\hline
		Sample & $\rho_0~(\Msun\pc^{-3})$ &  $z_\odot~(\pc)$ & $W_\odot~(\kmsec)$   \\
		\hline
		%\fcolorbox{black}{green}{\rule{0pt}{4pt}\rule{4pt}{0pt}}\quad
        S1 & $0.119^{+0.015}_{-0.012}$ & $15.29^{+2.24}_{-2.16}$ & $7.19^{+0.18}_{-0.18}$ \\
		\hline
        %\fcolorbox{black}{blue}{\rule{0pt}{4pt}\rule{4pt}{0pt}}\quad
        S2 & $0.096^{+0.021}_{-0.019}$ & $14.93^{+4.91}_{-4.43}$ & $7.17^{+0.28}_{-0.28}$ \\
		\hline
        %\fcolorbox{black}{red}{\rule{0pt}{4pt}\rule{4pt}{0pt}}\quad
        S3 & $0.118^{+0.013}_{-0.012}$ & $16.86^{+2.91}_{-2.78}$ & $6.9^{+0.24}_{-0.24}$ \\
		\hline
        S4 & $0.068^{+0.004}_{-0.004}$ & $13.39^{+1.79}_{-1.80}$ & $7.18^{+0.13}_{-0.13}$ \\
		\hline
	\end{tabular}
\end{table}

\begin{figure}
	\includegraphics[width=\columnwidth]{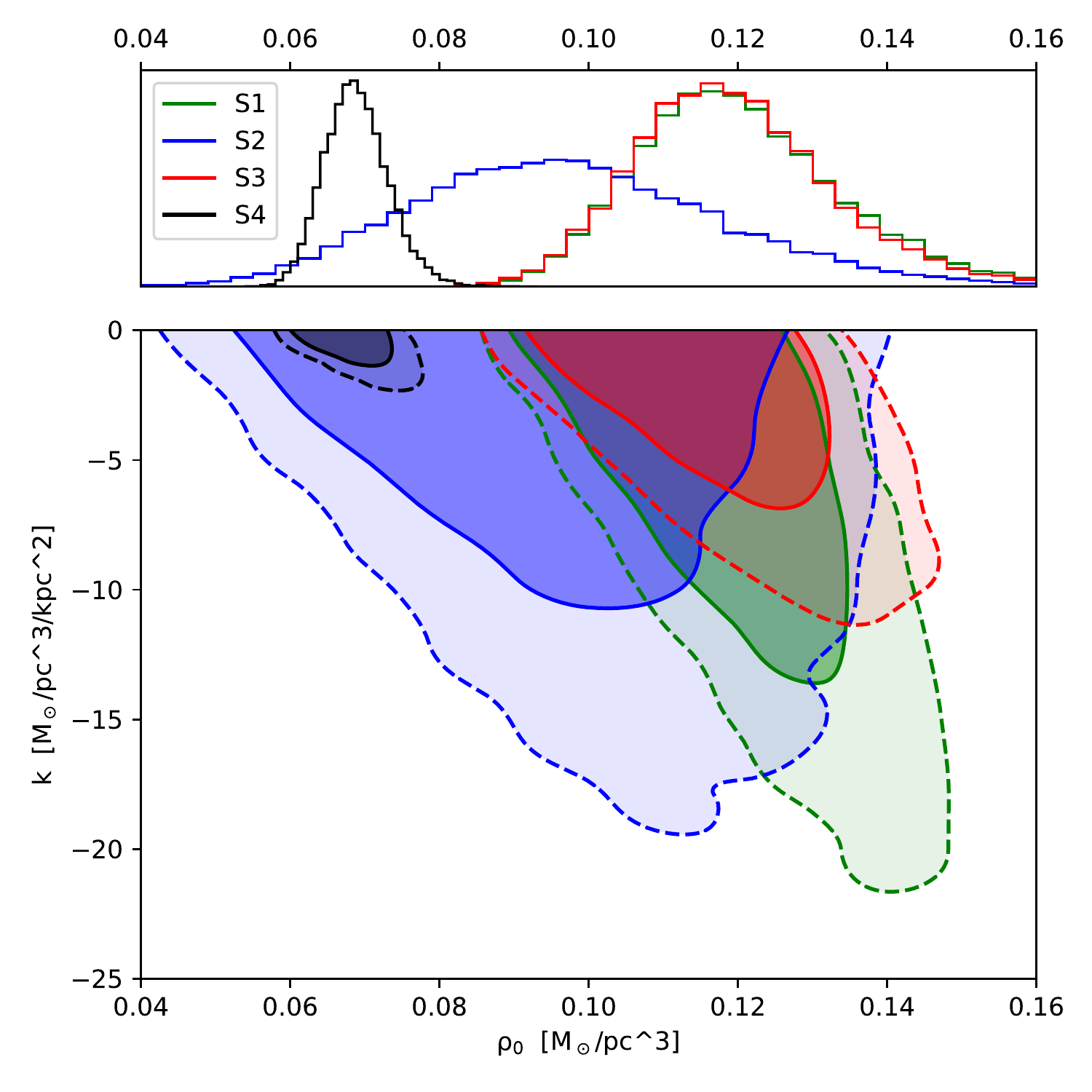}
    \caption{The lower panel shows the posterior probability for samples S1--S4 (in order green, blue, red, black), in the space of population parameters $\rho_0$ and $k$, marginalized over all other parameters. The region enclosed by the solid (dashed) line contains 68 per cent (90 per cent) of the total posterior density. The upper panel shows a histogram over $\rho_0$ only, where also $k$ is marginalized over. For visibility, the bin size of sample S4 half that of the other samples.}
    \label{fig:posteriors}
\end{figure}

The median posterior values for $\rho_0$, $z_\odot$ and $W_\odot$ are presented in Table~\ref{tab:results}. In Fig.~\ref{fig:posteriors}, the posterior distributions are shown in the ($\rho_0,k)$ space. The posteriors over $z_\odot$ and $W_\odot$ are visible in Fig.~\ref{fig:z0_W0}. The posterior distributions of all population parameters are visible in Appendix~\ref{app:plots}.

All samples suffer to some degree from model shortcomings and unforeseen biases. There are biases in the parallax measurements in TGAS \newchanges{of order $\sim 0.3$ mas}, coming from subtle differences in how the Gaia and Tycho surveys operate \citep{TGASrel}. The further the distance to a star the more severe such a bias will be, as the parallax significance decreases with distance. \newchanges{At a distance of 250 pc, a 0.3 mas parallax error corresponds to a distance error of 20 pc; this can produce a significant systematic error especially if such a bias is uniform over larger areas of the sky.} Another issue with going to large distances is the assumed parabolic shape of the density profile, which does not account for possible heavy tails at high $|z|$. Less severe is the fact that we do not account for dust extinction in our analysis; this is a good approximation for stars that are close by, because the Sun is situated in a neighbourhood almost devoid of dust, up to at least $100~\pc$ \citep{Capitanio17}. For these reasons, we deem the stellar samples that extend to smaller maximum distances to be more trustworthy.

In the stellar samples we have used, there is some overlap in terms of what stars are included. For this reason we cannot present a result of the different posteriors in unison, as we would be using the same data multiple times. We present our end result in terms of the posterior of sample S1, listed in Table~\ref{tab:samples}. We do not consider sample S4 reliable; it is included to illustrate the limitations of our model, method and data set. Other samples serve as a testament to the method and the result's viability.

Disregarding sample S4, the results of the different samples are consistent. As seen in Fig.~\ref{fig:posteriors}, samples S1 and S3 agree very well in terms of the inferred vales for $\rho_0$ and $k$. Sample S2 has a much wider posterior density; it has fewer stars and a lower maximal observed distance, giving less leverage in terms of how quickly the stellar number density decreases with distance from the plane. Samples S1--S3 all have a trail of points extending diagonally into more negative values of $k$ and higher density in the plane $\rho_0$. In other words, the data does not very well constrain how concentrated the matter distribution is to the plane. \changes{In sample S1, both the stellar number density and the dynamical matter density fall off to half of their respective maximum values at around $160~\pc$ distance from the plane. This is also the approximate extent of the stellar sample, as given by the cut in observed parallax. The inferred posterior value is thus insensitive to the tails of these distributions at distances above $\sim 160~\pc$.} Going to further distance does of course better constrain how quickly the density decreases with $z$: sample S4 has the tightest constraint to the value of $k$. However, at distances as far as $200~\pc$ from the plane, higher order terms, which are not included in this analysis, will likely start to significantly affect the tails of the density distribution.

\begin{figure}	\includegraphics[width=\columnwidth]{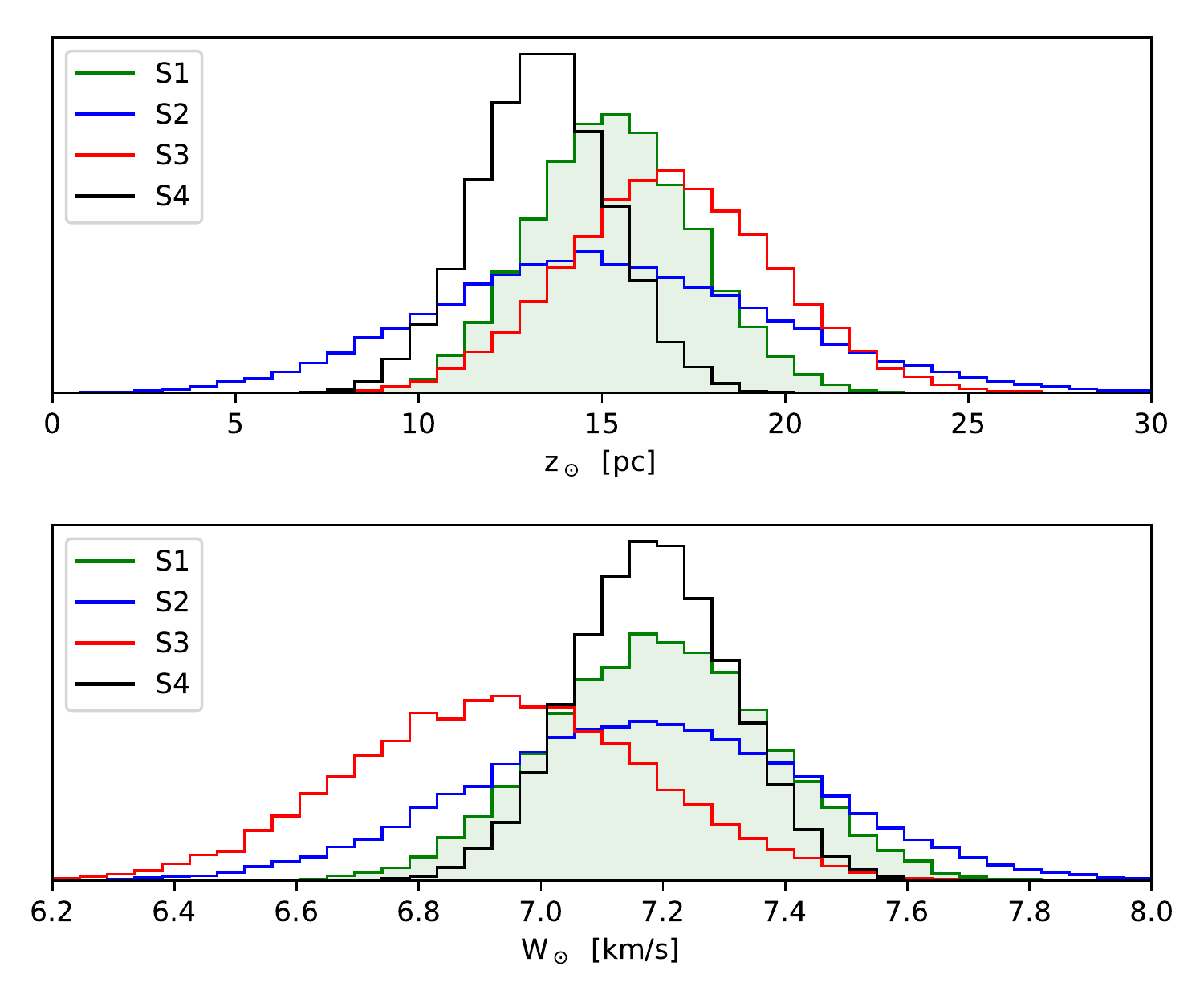}
    \caption{The posterior probability of population parameters $z_\odot$ (upper panel) and $W_\odot$ (lower panel), marginalized over all other parameters, for samples S1--S4 (in order green, blue, red, black). The area under the histogram of sample S1 is shaded green to guide the eye.}
    \label{fig:z0_W0}
\end{figure}

\begin{figure}
	\includegraphics[width=\columnwidth]{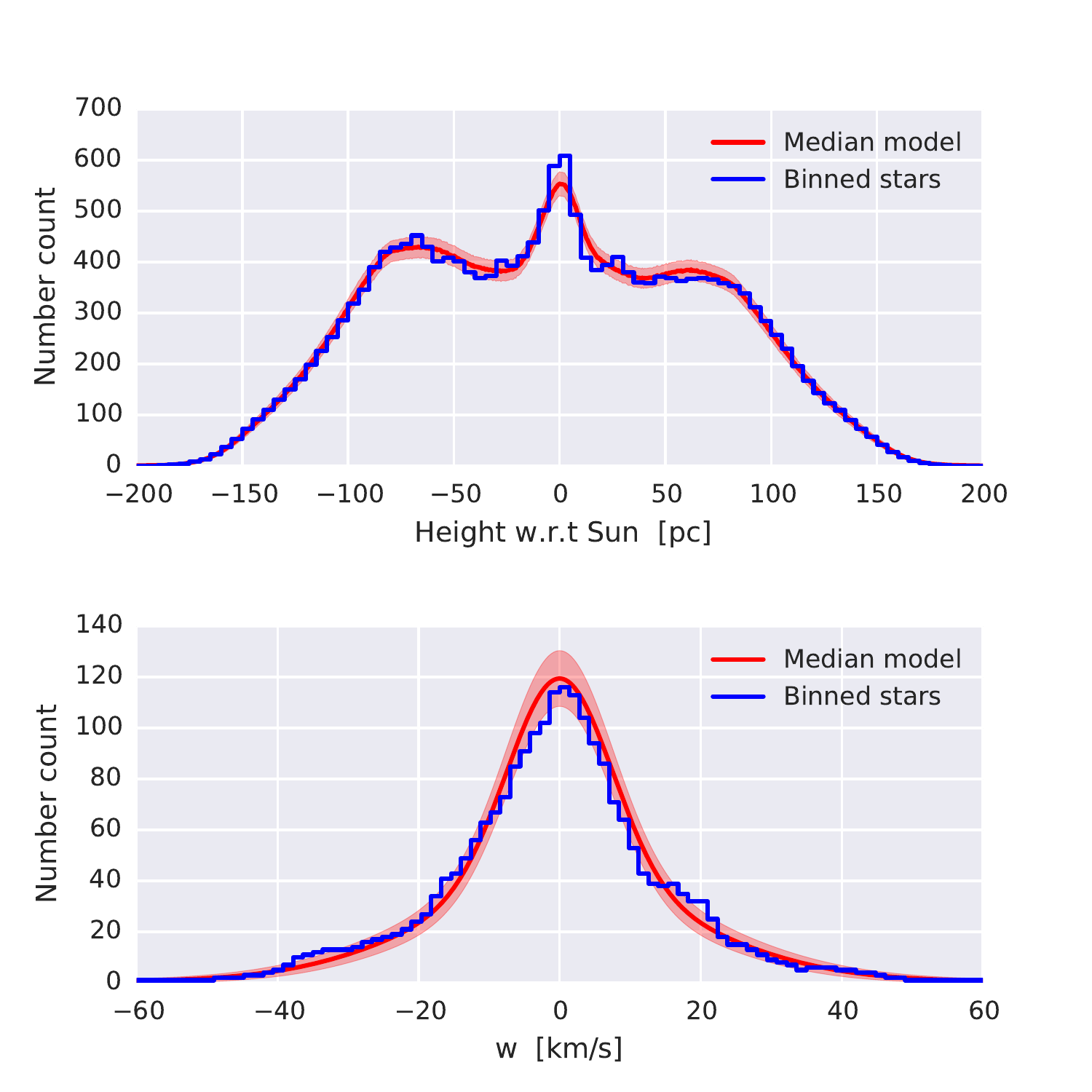}
    \caption{Comparison between model and inferred stellar parameters for sample S1. The model, represented by red graphs, have population parameters set to their respective posterior median values. The histogram over the stars, represented by blue graphs, are given by the inferred stellar parameters given the population model, i.e. $\text{Pr}(\bldpsi | \bldPsi_\text{med.},\bd,S)$. The upper panel shows the stars' height with respect to the Sun, where the model curve is proportional to $A_\text{eff}\tilde{\nu}$, normalized to account for the bin size and number of stars in the sample. The lower panel shows the velocity $w$ for a subset of stars with $|b|<5^\circ$, where the model is proportional to the velocity distribution in the plane, $\tilde{f}(0,w)$, normalized to the correct number of stars. The extent of the shaded red regions corresponds to a Poisson noise, whose width is given by the square root of the model curve.}
    \label{fig:fit}
\end{figure}

Fig.~\ref{fig:fit} illustrates how well the model describes the distribution of stars of sample S1, for a specific set of population parameters. The upper panel of this figure illustrates the number of stars of a certain height with respect to the Sun. The fit looks good, although there are some minor detractions from the model around the center peak. It is difficult to say the exact reason for this but possible reasons could be some local substructure, a selection effect not accounted for, that our density profile is not sophisticated enough, etc. In the lower panel, we show the velocity distribution on the Galactic plane $\tilde{f}(0,w)$ (red line) and the histogram of the $w$ distribution for stars with $|b|<5^\circ$ (blue line). There are some small but clear substructures in this velocity distribution, such as over-densities at $w\simeq 20~\kmsec$ and $w\simeq -35~\kmsec$, and a slight skewness of the distribution.

\section{Discussion and conclusions}\label{sect:concl}

In this work, we have estimated the total dynamical density of the Galaxy in the solar neighbourhood $\rho_0$, the Sun's height from the Galactic plane $z_\odot$, and vertical velocity $W_\odot$ using astrometric data from the TGAS catalogue, with full treatment of astrometric errors for each star. We have used different stellar samples, and for the one we deem most trustworthy we measured $\rho_0 = 0.119^{+0.015}_{-0.012}~\Msun\pc^{-3}$, $z_\odot=15.29^{+2.24}_{-2.16}~\pc$, and $W_\odot=7.19^{+0.18}_{-0.18}~\kmsec$. \newchanges{We obtain an excellent consistency of the results for $z_\odot$ and $W_\odot$ in all samples. In terms of $\rho_0$, the results are $\rho_0 = 0.096^{+0.021}_{-0.019}~\Msun\pc^{-3}$ for the second sample, $\rho_0 = 0.118^{+0.013}_{-0.012}~\Msun\pc^{-3}$ for the third sample, and $\rho_0 = 0.068^{+0.004}_{-0.004}~\Msun\pc^{-3}$ for the fourth sample. These results are in statistical agreement, with the exception of the fourth sample. We consider the fourth sample to be least trustworthy, especially because the simple form of the gravitational potential is expected to be invalid at such high distances from the plane. The second sample has a very poor leverage on the matter density due to having few stars and staying very close in distance. The first and third samples have about equal leverage on $\rho_0$ and agree very well.}

Our analysis was directly inspired by the works of \cite{Creze1998} and \cite{HolmbergFlynn2000}, which used astrometry from the Hipparcos catalogue to estimate the total dynamical density in the solar neighbourhood. Our result for $\rho_0$ is well in line with the result of \cite{HolmbergFlynn2000}, while \cite{Creze1998} estimated $\rho_0$ to be significantly lower. An important difference between our analysis and these works is that our fitting algorithm accounts for measurement uncertainties on all individual stars, while their estimates come from fits on binned data \citep[although in][they approximated the influence of such errors by Monte-Carlo sampling of mock data catalogues]{HolmbergFlynn2000}. Another difference is that, while Hipparcos was a complete catalogue of stars, TGAS is very incomplete in many respects and modelling the selection function of this catalogue is complex.   
For this reason \cite{Creze1998} and \cite{HolmbergFlynn2000} could use bright A and F stars, which have a lower velocity dispersions and whose number density falls off faster with height from the Galactic plane. We are forced to use fainter stars, since TGAS has poor completeness for bright objects. \newchanges{Fainter stars are typically kinematically warmer and therefore more affected by the large-scale structure of the Galaxy, but are on the other hand also better phase--mixed than brighter stars.} The stellar samples that we consider reliable (which extend no further than $\sim 200~\pc$) cannot infer much about the shape of the density distribution. The curvature $k$ more or less fills the prior space set by the population parameter prior. \changes{In Appendix~\ref{app:toydata}, we apply the same method of inference on mock data and get an equally poor constraint on $k$, making it clear that the quality of the data is inadequate in terms of strongly constraining this parameter. Adopting a different matter distribution with heavier tails (such as $\text{sech}^2$ or similar) would give very similar results to the central density $\rho_0$, degenerate with a small adjustment to the population parameter prior.} The sample that contains the furthest stars used in our work has a stronger inference on the dynamical matter density and its shape. Unfortunately, we cannot take this result very seriously, as we suspect it to be biased by systematic errors in parallax; moreover, going to these heights would necessitate a more sophisticated density profile, that accounts for the behavior of its tails. \cite{HolmbergFlynn2000} associate the difference of their results on $\rho_0$ as compared to \cite{Creze1998} to the more sophisticated potential/density model that they use. Interestingly, we find results more in line with those of \cite{HolmbergFlynn2000}, even if we impose a constant density as used in \cite{Creze1998}. The density of visible matter in the solar neighbourhood is estimated to be of the order of $0.09\pm0.01~\Msun\pc^{-3}$ \citep{Flynn2006}, meaning that we can rule out large density contributions by dark matter, \changes{such as from a dark disc. A dark disk can be formed by accretion of satellites \citep{Read2008} or by dissipation through self--interactions of a sub--dominant dark matter component \citep{2013PDU.....2..139F}. In the former scenario of satellite accretion, the dark disk would be warmer and thicker, probably extending to more than 200 pc from the disk. Our result is not sensitive to the shape of the density curve at such high distances. In the latter scenario of a dark disk formed from dark matter dissipation, the disk would be thinner and therefore in principle detectable using this method. Detection could still be problematic though, as the dark disc could be locally out of equilibrium \citep{2016ApJ...824..116K}.}

\changes{Our analysis is based on the assumption that, for most of the stars in our samples, the vertical energy $E_z$ is an integral of motion, i.e. stars have decoupled horizontal and vertical motions. \cite{Garbari2011} have shown that in cosmological simulations of the Milky Way, stars that reach a distance from the Galactic plane of about 500~pc or more, have significant coupling in their motions. In the case of the potential and distribution function given by the median parameters as inferred from sample S1, only about 4 per cent of the stars on the Galactic plane have sufficient vertical velocities ($|v_z| \gtrsim 40~\kmsec$) to reach 500~pc height or more. It is not trivial to say how a coupling of vertical and horizontal motion for stars at such distances would affect their velocities when passing through the plane, which in turn would bias our result. In order to test how severe such a bias could be, we assume an extreme case, where high--velocity stars are maximally biasing our result by only contributing to the number density close to the plane. A mid--plane number density that is over--estimated by 4 per cent would mimic the effects of a greater dynamical matter density, producing a bias this is roughly equal to the statistical uncertainty of sample S1. However, such a conspiracy seems extremely contrived and unlikely, which is why we expect this bias to be small compared to other errors.}

Our results on the height of the Sun from the Galactic plane are reasonable and well in line with some older estimates \citep[e.g., from the NIR stars distribution in the Galactic plane observed by COBE/DIRBE,][]{Binney1997} and very recent ones from the vertical distribution of pulsars in the Galaxy \citep{Yao2017}. Some estimates obtained from star counts are higher, for example from the SDSS survey \citep[$z_\odot=25\pm 5~\pc$ in][]{Juric2008}. Our estimate for the Sun's vertical velocity agrees well with most studies on the local kinematics of the Milky Way \citep[e.g.][]{Schonrich2010}.

Analyzing the distribution of vertical velocities of stars near the Galactic plane, seen in Fig.~\ref{fig:fit}, we observe the presence of small but significant substructures in the form of over-densities, the two most significant traveling at approximately $-35~\kmsec$ and $20~\kmsec$. We also notice a certain skewness in the whole distribution. These features do not seem to be explained by known open clusters. They could be interpreted as remnants of small merger events or as excitations produced by an outside system \citep{Widrow2014}, while a plane-symmetric perturbation to the disc, like the bar or the spiral arms, could not create such features \citep{Monari2016}. \newchanges{We caution the reader that our modeling of the dynamics of the solar neighborhood assumes axisymmetry and plane symmetry}. The proper modelling of such perturbations will represent a challenge for the interpretation of the new Gaia data \citep{Banik}.

In conclusion, we hope to repeat this type of analysis with the next Gaia data release (DR2). With DR2, the astrometric precision will be much improved and the catalogue will be practically complete down to some apparent magnitude. The analysis will be simpler in many ways, since selection effects will be much less severe than what is currently the case for TGAS. DR2 will also contain spectroscopically determined line-of-sight velocities, making it possible to include velocity information also for stars far from the Galactic plane. With these improvements, we are optimistic about being able to say something more definitive concerning the shape of the dynamical density distribution in the solar neighbourhood. This is very important, as it can constrain dark matter particle candidates, and give information about the dynamics and formation of the Milky Way.

\section*{Acknowledgements}

We would like to thank Daniel Mortlock, with whom we have had interesting and insightful discussions about statistics in general and Bayesian hierarchical modelling in particular. We would also like to thank Olivier Bienaym\'{e}, Benoit Famaey, and Douglas Spolyar for useful discussions, \changes{and our anonymous referees for a detailed and constructive peer review.}

%%%%%%%%%%%%%%%%%%%%%%%%%%%%%%%%%%%%%%%%%%%%%%%%%%



\bibliographystyle{mnras}
\bibliography{locdenbib} % if your bibtex file is called example.bib

\begin{thebibliography}{}
\makeatletter
\relax
\def\mn@urlcharsother{\let\do\@makeother \do\$\do\&\do\#\do\^\do\_\do\%\do\~}
\def\mn@doi{\begingroup\mn@urlcharsother \@ifnextchar [ {\mn@doi@}
  {\mn@doi@[]}}
\def\mn@doi@[#1]#2{\def\@tempa{#1}\ifx\@tempa\@empty \href
  {http://dx.doi.org/#2} {doi:#2}\else \href {http://dx.doi.org/#2} {#1}\fi
  \endgroup}
\def\mn@eprint#1#2{\mn@eprint@#1:#2::\@nil}
\def\mn@eprint@arXiv#1{\href {http://arxiv.org/abs/#1} {{\tt arXiv:#1}}}
\def\mn@eprint@dblp#1{\href {http://dblp.uni-trier.de/rec/bibtex/#1.xml}
  {dblp:#1}}
\def\mn@eprint@#1:#2:#3:#4\@nil{\def\@tempa {#1}\def\@tempb {#2}\def\@tempc
  {#3}\ifx \@tempc \@empty \let \@tempc \@tempb \let \@tempb \@tempa \fi \ifx
  \@tempb \@empty \def\@tempb {arXiv}\fi \@ifundefined
  {mn@eprint@\@tempb}{\@tempb:\@tempc}{\expandafter \expandafter \csname
  mn@eprint@\@tempb\endcsname \expandafter{\@tempc}}}

\bibitem[\protect\citeauthoryear{{Banik}, {Widrow}  \& {Dodelson}}{{Banik}
  et~al.}{2017}]{Banik}
{Banik} N.,  {Widrow} L.~M.,   {Dodelson} S.,  2017, \mn@doi [\mnras]
  {10.1093/mnras/stw2603}, \href
  {http://adsabs.harvard.edu/abs/2017MNRAS.464.3775B} {464, 3775}

\bibitem[\protect\citeauthoryear{{Binney} \& {Merrifield}}{{Binney} \&
  {Merrifield}}{1998}]{BinneyMerrifield}
{Binney} J.,  {Merrifield} M.,  1998, {Galactic Astronomy}

\bibitem[\protect\citeauthoryear{{Binney} \& {Tremaine}}{{Binney} \&
  {Tremaine}}{2008}]{BT2008}
{Binney} J.,  {Tremaine} S.,  2008, {Galactic Dynamics: Second Edition}.
Princeton University Press

\bibitem[\protect\citeauthoryear{{Binney}, {Gerhard}  \& {Spergel}}{{Binney}
  et~al.}{1997}]{Binney1997}
{Binney} J.,  {Gerhard} O.,   {Spergel} D.,  1997, \mn@doi [\mnras]
  {10.1093/mnras/288.2.365}, \href
  {http://cdsads.u-strasbg.fr/abs/1997MNRAS.288..365B} {288, 365}

\bibitem[\protect\citeauthoryear{{Capitanio}, {Lallement}, {Vergely},
  {Elyajouri}  \& {Monreal-Ibero}}{{Capitanio} et~al.}{2017}]{Capitanio17}
{Capitanio} L.,  {Lallement} R.,  {Vergely} J.~L.,  {Elyajouri} M.,
  {Monreal-Ibero} A.,  2017, \mn@doi [\aap] {10.1051/0004-6361/201730831},
  \href {http://adsabs.harvard.edu/abs/2017A%26A...606A..65C} {606, A65}

\bibitem[\protect\citeauthoryear{{Cr\'ez\'e}, {Chereul}, {Bienayme}  \&
  {Pichon}}{{Cr\'ez\'e} et~al.}{1998}]{Creze1998}
{Cr\'ez\'e} M.,  {Chereul} E.,  {Bienayme} O.,   {Pichon} C.,  1998, \aap,
  \href {http://cdsads.u-strasbg.fr/abs/1998A%26A...329..920C} {329, 920}

\bibitem[\protect\citeauthoryear{{Dalton} et~al.,}{{Dalton}
  et~al.}{2014}]{WEAVE}
{Dalton} G.,  et~al., 2014, in Ground-based and Airborne Instrumentation for
  Astronomy V. p. 91470L (\mn@eprint {arXiv} {1412.0843}),
  \mn@doi{10.1117/12.2055132}

\bibitem[\protect\citeauthoryear{{Fan}, {Katz}, {Randall}  \& {Reece}}{{Fan}
  et~al.}{2013}]{2013PDU.....2..139F}
{Fan} J.,  {Katz} A.,  {Randall} L.,   {Reece} M.,  2013, \mn@doi [Physics of
  the Dark Universe] {10.1016/j.dark.2013.07.001}, \href
  {http://adsabs.harvard.edu/abs/2013PDU.....2..139F} {2, 139}

\bibitem[\protect\citeauthoryear{{Flynn}, {Holmberg}, {Portinari}, {Fuchs}  \&
  {Jahrei{\ss}}}{{Flynn} et~al.}{2006}]{Flynn2006}
{Flynn} C.,  {Holmberg} J.,  {Portinari} L.,  {Fuchs} B.,   {Jahrei{\ss}} H.,
  2006, \mn@doi [\mnras] {10.1111/j.1365-2966.2006.10911.x}, \href
  {http://adsabs.harvard.edu/abs/2006MNRAS.372.1149F} {372, 1149}

\bibitem[\protect\citeauthoryear{Foreman-Mackey}{Foreman-Mackey}{2016}]{corner}
Foreman-Mackey D.,  2016, \mn@doi [The Journal of Open Source Software]
  {10.21105/joss.00024}, 24

\bibitem[\protect\citeauthoryear{{Gaia Collaboration} et~al.,}{{Gaia
  Collaboration} et~al.}{2016a}]{Gaia}
{Gaia Collaboration} et~al., 2016a, \mn@doi [\aap]
  {10.1051/0004-6361/201629272}, \href
  {http://adsabs.harvard.edu/abs/2016A%26A...595A...1G} {595, A1}

\bibitem[\protect\citeauthoryear{{Gaia Collaboration} et~al.,}{{Gaia
  Collaboration} et~al.}{2016b}]{GaiaDR1}
{Gaia Collaboration} et~al., 2016b, \mn@doi [\aap]
  {10.1051/0004-6361/201629512}, \href
  {http://adsabs.harvard.edu/abs/2016A%26A...595A...2G} {595, A2}

\bibitem[\protect\citeauthoryear{{Garbari}, {Read}  \& {Lake}}{{Garbari}
  et~al.}{2011}]{Garbari2011}
{Garbari} S.,  {Read} J.~I.,   {Lake} G.,  2011, \mn@doi [\mnras]
  {10.1111/j.1365-2966.2011.19206.x}, \href
  {http://adsabs.harvard.edu/abs/2011MNRAS.416.2318G} {416, 2318}

\bibitem[\protect\citeauthoryear{{Gelman}, {Carlin}, {Stern}, {Dunson},
  {Vehtari}  \& {Rubin}}{{Gelman} et~al.}{2013}]{BayesianDataAnalysis}
{Gelman} A.,  {Carlin} J.,  {Stern} H.,  {Dunson} D.,  {Vehtari} A.,   {Rubin}
  D.,  2013, {Bayesian Data Analysis: Third Edition}

\bibitem[\protect\citeauthoryear{{H{\o}g} et~al.,}{{H{\o}g}
  et~al.}{2000}]{Tycho2}
{H{\o}g} E.,  et~al., 2000, \aap, \href
  {http://adsabs.harvard.edu/abs/2000A%26A...355L..27H} {355, L27}

\bibitem[\protect\citeauthoryear{{Holmberg} \& {Flynn}}{{Holmberg} \&
  {Flynn}}{2000}]{HolmbergFlynn2000}
{Holmberg} J.,  {Flynn} C.,  2000, \mn@doi [\mnras]
  {10.1046/j.1365-8711.2000.02905.x}, \href
  {http://cdsads.u-strasbg.fr/abs/2000MNRAS.313..209H} {313, 209}

\bibitem[\protect\citeauthoryear{{Juri{\'c}} et~al.,}{{Juri{\'c}}
  et~al.}{2008}]{Juric2008}
{Juri{\'c}} M.,  et~al., 2008, \mn@doi [\apj] {10.1086/523619}, \href
  {http://adsabs.harvard.edu/abs/2008ApJ...673..864J} {673, 864}

\bibitem[\protect\citeauthoryear{{Kapteyn}}{{Kapteyn}}{1922}]{Kapteyn1922}
{Kapteyn} J.~C.,  1922, \mn@doi [\apj] {10.1086/142670}, \href
  {http://cdsads.u-strasbg.fr/abs/1922ApJ....55..302K} {55, 302}

\bibitem[\protect\citeauthoryear{{Kramer} \& {Randall}}{{Kramer} \&
  {Randall}}{2016}]{2016ApJ...824..116K}
{Kramer} E.~D.,  {Randall} L.,  2016, \mn@doi [\apj]
  {10.3847/0004-637X/824/2/116}, \href
  {http://adsabs.harvard.edu/abs/2016ApJ...824..116K} {824, 116}

\bibitem[\protect\citeauthoryear{{Kuijken} \& {Gilmore}}{{Kuijken} \&
  {Gilmore}}{1989a}]{KuijkenGilmore1989a}
{Kuijken} K.,  {Gilmore} G.,  1989a, \mn@doi [\mnras]
  {10.1093/mnras/239.2.571}, \href
  {http://cdsads.u-strasbg.fr/abs/1989MNRAS.239..571K} {239, 571}

\bibitem[\protect\citeauthoryear{{Kuijken} \& {Gilmore}}{{Kuijken} \&
  {Gilmore}}{1989b}]{KuijkenGilmore1989b}
{Kuijken} K.,  {Gilmore} G.,  1989b, \mn@doi [\mnras]
  {10.1093/mnras/239.2.605}, \href
  {http://cdsads.u-strasbg.fr/abs/1989MNRAS.239..605K} {239, 605}

\bibitem[\protect\citeauthoryear{{Kuijken} \& {Gilmore}}{{Kuijken} \&
  {Gilmore}}{1989c}]{KuijkenGilmore1989c}
{Kuijken} K.,  {Gilmore} G.,  1989c, \mn@doi [\mnras]
  {10.1093/mnras/239.2.651}, \href
  {http://cdsads.u-strasbg.fr/abs/1989MNRAS.239..651K} {239, 651}

\bibitem[\protect\citeauthoryear{{Kuijken} \& {Gilmore}}{{Kuijken} \&
  {Gilmore}}{1991}]{KuijkenGilmore1991}
{Kuijken} K.,  {Gilmore} G.,  1991, \mn@doi [\apjl] {10.1086/185920}, \href
  {http://cdsads.u-strasbg.fr/abs/1991ApJ...367L...9K} {367, L9}

\bibitem[\protect\citeauthoryear{{Lindegren} et~al.,}{{Lindegren}
  et~al.}{2016}]{TGASrel}
{Lindegren} L.,  et~al., 2016, \mn@doi [\aap] {10.1051/0004-6361/201628714},
  \href {http://adsabs.harvard.edu/abs/2016A%26A...595A...4L} {595, A4}

\bibitem[\protect\citeauthoryear{{McMillan} \& {Binney}}{{McMillan} \&
  {Binney}}{2013}]{McMillanBinney2013}
{McMillan} P.~J.,  {Binney} J.~J.,  2013, \mn@doi [\mnras]
  {10.1093/mnras/stt814}, \href
  {http://adsabs.harvard.edu/abs/2013MNRAS.433.1411M} {433, 1411}

\bibitem[\protect\citeauthoryear{{Michalik}, {Lindegren}  \&
  {Hobbs}}{{Michalik} et~al.}{2015}]{TGAS}
{Michalik} D.,  {Lindegren} L.,   {Hobbs} D.,  2015, \mn@doi [\aap]
  {10.1051/0004-6361/201425310}, \href
  {http://adsabs.harvard.edu/abs/2015A%26A...574A.115M} {574, A115}

\bibitem[\protect\citeauthoryear{{Monari}, {Famaey}  \& {Siebert}}{{Monari}
  et~al.}{2016}]{Monari2016}
{Monari} G.,  {Famaey} B.,   {Siebert} A.,  2016, \mn@doi [\mnras]
  {10.1093/mnras/stw171}, \href
  {http://cdsads.u-strasbg.fr/abs/2016MNRAS.457.2569M} {457, 2569}

\bibitem[\protect\citeauthoryear{{Oort}}{{Oort}}{1932}]{Oort1932}
{Oort} J.~H.,  1932, \bain, \href
  {http://cdsads.u-strasbg.fr/abs/1932BAN.....6..249O} {6, 249}

\bibitem[\protect\citeauthoryear{{Perryman} et~al.,}{{Perryman}
  et~al.}{1997}]{Hipparcos}
{Perryman} M.~A.~C.,  et~al., 1997, \aap, \href
  {http://adsabs.harvard.edu/abs/1997A%26A...323L..49P} {323, L49}

\bibitem[\protect\citeauthoryear{Poincar\'e}{Poincar\'e}{1906}]{Poincare1906}
Poincar\'e H.,  1906, Bulletin de la soci\'et\'e astronomique de France, 20,
  153

\bibitem[\protect\citeauthoryear{{Read}}{{Read}}{2014}]{Read2014}
{Read} J.~I.,  2014, \mn@doi [Journal of Physics G Nuclear Physics]
  {10.1088/0954-3899/41/6/063101}, \href
  {http://cdsads.u-strasbg.fr/abs/2014JPhG...41f3101R} {41, 063101}

\bibitem[\protect\citeauthoryear{{Read}, {Lake}, {Agertz}  \&
  {Debattista}}{{Read} et~al.}{2008}]{Read2008}
{Read} J.~I.,  {Lake} G.,  {Agertz} O.,   {Debattista} V.~P.,  2008, \mn@doi
  [\mnras] {10.1111/j.1365-2966.2008.13643.x}, \href
  {http://adsabs.harvard.edu/abs/2008MNRAS.389.1041R} {389, 1041}

\bibitem[\protect\citeauthoryear{{Sch{\"o}nrich}, {Binney}  \&
  {Dehnen}}{{Sch{\"o}nrich} et~al.}{2010}]{Schonrich2010}
{Sch{\"o}nrich} R.,  {Binney} J.,   {Dehnen} W.,  2010, \mn@doi [\mnras]
  {10.1111/j.1365-2966.2010.16253.x}, \href
  {http://adsabs.harvard.edu/abs/2010MNRAS.403.1829S} {403, 1829}

\bibitem[\protect\citeauthoryear{{Widrow}, {Barber}, {Chequers}  \&
  {Cheng}}{{Widrow} et~al.}{2014}]{Widrow2014}
{Widrow} L.~M.,  {Barber} J.,  {Chequers} M.~H.,   {Cheng} E.,  2014, \mn@doi
  [\mnras] {10.1093/mnras/stu396}, \href
  {http://adsabs.harvard.edu/abs/2014MNRAS.440.1971W} {440, 1971}

\bibitem[\protect\citeauthoryear{{Yao}, {Manchester}  \& {Wang}}{{Yao}
  et~al.}{2017}]{Yao2017}
{Yao} J.~M.,  {Manchester} R.~N.,   {Wang} N.,  2017, \mn@doi [\mnras]
  {10.1093/mnras/stx729}, \href
  {http://adsabs.harvard.edu/abs/2017MNRAS.468.3289Y} {468, 3289}

\bibitem[\protect\citeauthoryear{{de Jong} et~al.,}{{de Jong}
  et~al.}{2012}]{4MOST}
{de Jong} R.~S.,  et~al., 2012, in Ground-based and Airborne Instrumentation
  for Astronomy IV. p. 84460T (\mn@eprint {arXiv} {1206.6885}),
  \mn@doi{10.1117/12.926239}

\bibitem[\protect\citeauthoryear{{van der Kruit} \& {van Berkel}}{{van der
  Kruit} \& {van Berkel}}{2000}]{KapteynLegacy}
{van der Kruit} P.~C.,  {van Berkel} K.,  eds, 2000, {The legacy of J. C.
  Kapteyn. Studies on Kapteyn and the development of modern astronomy}
  Astrophysics and Space Science Library Vol. 246,
  \mn@doi{10.1007/978-94-010-9864-9.
}

\makeatother
\end{thebibliography}

%\begin{thebibliography}{99}
%\end{thebibliography}

%%%%%%%%%%%%%%%%%%%%%%%%%%%%%%%%%%%%%%%%%%%%%%%%%%

%%%%%%%%%%%%%%%%% APPENDICES %%%%%%%%%%%%%%%%%%%%%

\appendix

\section{Inference on mock data}\label{app:toydata}

We demonstrate our statistical method on samples of mock data, generated from a population model. The mock data samples are generated by rejection sampling, which works as follows. A star is drawn at random from the population model. An observational uncertainty is assigned to it, as well as observational quantities with errors included. It is then subjected to the selection function, which determines whether or not it is included in the sample. This is repeated until the desired number of stars are included in the sample.

The population parameters of the mock data samples are taken to be
%$\rho_0=0.12~\Msun\pc^{-3}$; $k=-6~\Msun\pc^{-3}\kpc^{-2}$; $\alpha = 0.5$; $\beta = 0.9$; $\sigma_{1,2,3}=\{7,18,40\}~\kmsec$; $z_0=0~\pc$; $W_\odot=7.2~\kmsec$.
\begin{equation}
\begin{split}
	&\rho_0 = 0.12~\Msun\pc^{-3},\\
    &k = -6~\Msun\pc^{-3}\kpc^{-2},\\
    &\alpha = 0.5,\\
    &\beta = 0.9,\\
    &\sigma_{1,2,3} = \{7,18,40\}~\kmsec,\\
    &z_0 = 0~\pc,\\
    &W_\odot = 7.2~\kmsec.
\end{split}
\end{equation}
Together with the luminosity function $F(M_V)$, this describes the population model, as detailed in \Sect{sect:pop_model}. From this model we can generate stars (either analytically or numerically through rejection sampling).

Parallax uncertainty is taken from the distribution of parallax uncertainties in TGAS as a function of apparent magnitude, called $g(\sigma_\varpi,m_V)$ in equation \eqref{eq:likelihood_int}, also visualized in Figure \ref{fig:parallax_error_percentiles}. Proper motion uncertainties and parallax--proper motion correlations are taken from the real data sample S1, drawn randomly from bins with a width of 0.5 mag in apparent magnitude $m_V$, where motion in directions parallel to the plane are also accounted for.

The selection effects come from cuts to observable quantities of our sample construction, as well as masking and completeness effects. We make mock data samples with cuts in observed parallax and observed absolute magnitude corresponding to those of sample S1, such that $M_V \in [3,5]$ and $I_\varpi \in [6.25,12.5]$. A star is not included in the sample if it falls in a region of the sky that is masked, as discussed in Section \ref{sec:masking}. Given that it falls in a non--masked region, it has a probability of being included that is equal to the completeness function, which is a function of angular position and apparent magnitude, as described in \Sect{sec:completeness}.

We construct two mock data samples, called M1 and M2, containing 20,000 stars respectively. The inference on the population parameters of these samples are performed in the same manner as for the real data samples, as described in \Sect{sec:sampling}.

\begin{figure}
\includegraphics[width=\columnwidth]{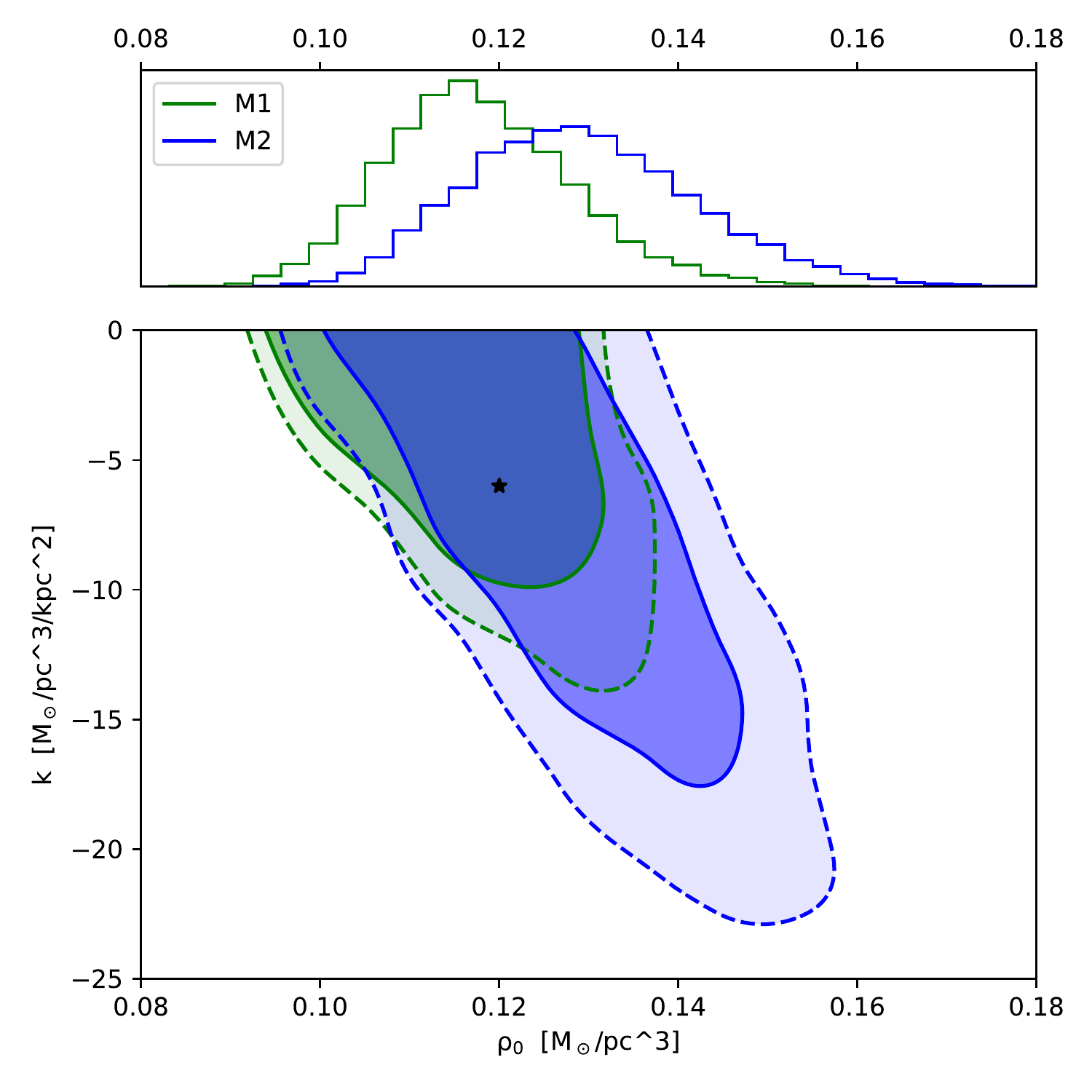}
    \caption{The lower panel shows the posterior probability for mock data samples M1 and M2 (green and blue), in the space of population parameters $\rho_0$ and $k$, marginalized over all other parameters. The region enclosed by the solid (dashed) line contains 68 per cent (90 per cent) of the total posterior density. The upper panel shows a histogram over $\rho_0$ only, where also $k$ is marginalized over. The true values of population parameters $\rho_0$ and $k$ of the mock data samples are marked with a black star.}
    \label{fig:toy_posteriors}
\end{figure}

Results for $\rho_0$ and $k$ are visible in Fig. \ref{fig:toy_posteriors}, where the true values are retrieved. The median posterior values of the density $\rho_0$ and the Sun's position and vertical velocity are listed in Table~\ref{tab:toy_results}. The true posterior values are retrieved within the 16th--84th percentiles of these population parameters, with the exception of $z_0$ for sample M1, which is slightly biased towards a higher value (the true value is found around the 7th percentile).

The shape of the density distribution is poorly constrained for these mock data samples, similar to the real sample S1. This demonstrates that the poor inference on the curvature $k$ is due to limitations in data quality (as opposed to the model being incorrect, for example).

\begin{table}
	\centering
	\caption{Posterior results of the dynamical matter density and the Sun's position and velocity with respect to the plane, presented as the median plus/minus its 16th and 84th percentiles, for mock data samples M1 and M2.}
    \label{tab:toy_results}
	\begin{tabular}{r || l l l}
		\hline
		Sample & $\rho_0~(\Msun\pc^{-3})$ &  $z_\odot~(\pc)$ & $W_\odot~(\kmsec)$   \\
		\hline
        M1 & $0.117^{+0.011}_{-0.009}$ & $2.60^{+1.69}_{-1.72}$ & $7.30^{+0.16}_{-0.16}$ \\
		\hline
        M2 & $0.129^{+0.013}_{-0.012}$ & $-0.84^{+1.75}_{-1.80}$ & $7.25^{+0.16}_{-0.16}$ \\
		\hline
	\end{tabular}
\end{table}

\section{Full population parameter posteriors}\label{app:plots}

\begin{figure*}
	\includegraphics[width=\linewidth]{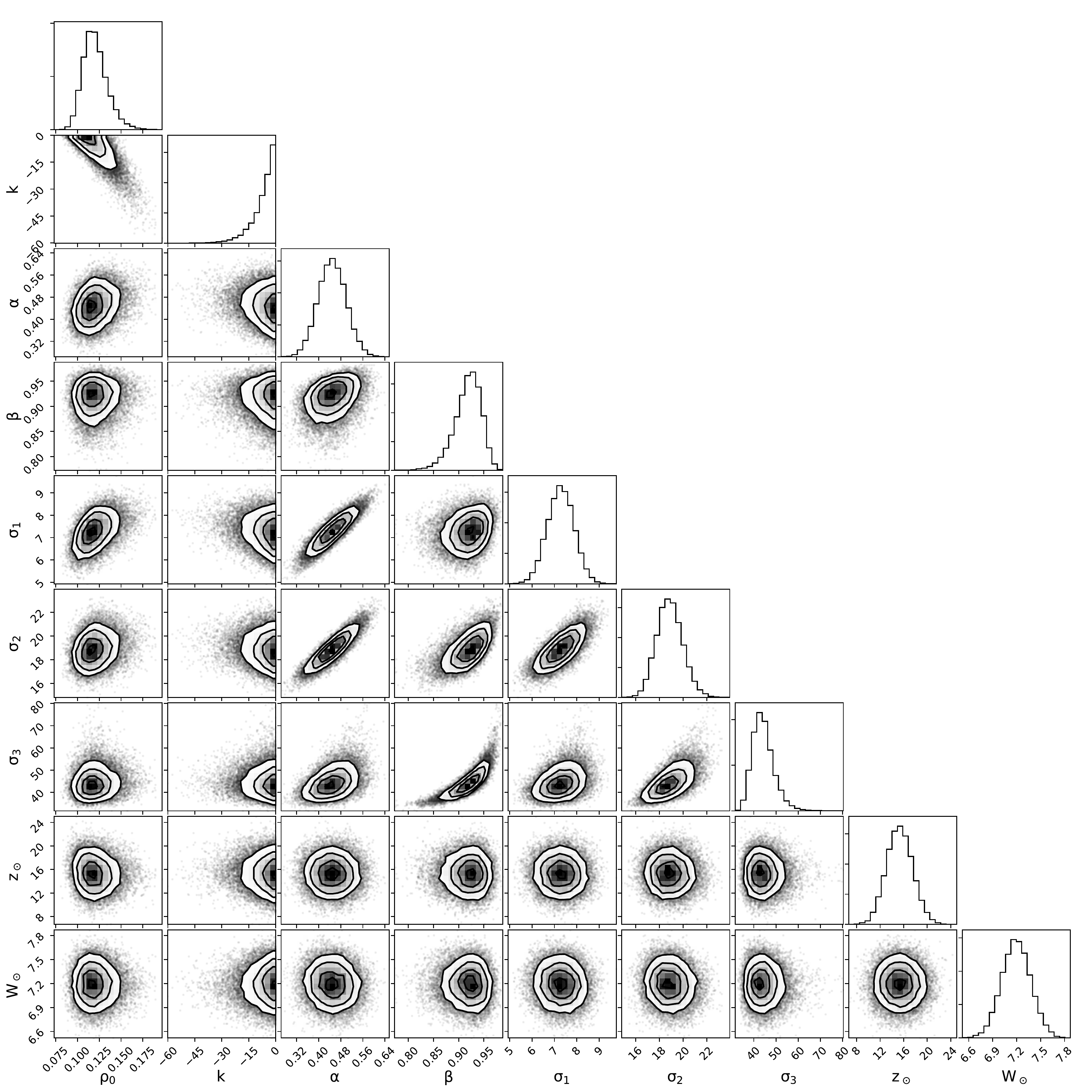}
    \caption{Population parameters posterior density of sample S1. The quantities have the following units: $\rho_0$ [$\Msun\pc^{-3}$]; $k$ [$\Msun\pc^{-3}\kpc^{-2}$]; $\alpha$ and $\beta$ [unitless]; $\sigma_{1,2,3}[\kmsec]$; $z_\odot[\pc]$; $W_\odot[\kmsec]$. The contours correspond to 0.5-, 1-, 1.5- and 2-$\sigma$ levels.}
    \label{fig:cornerS1}
\end{figure*}

\begin{figure*}
	\includegraphics[width=\linewidth]{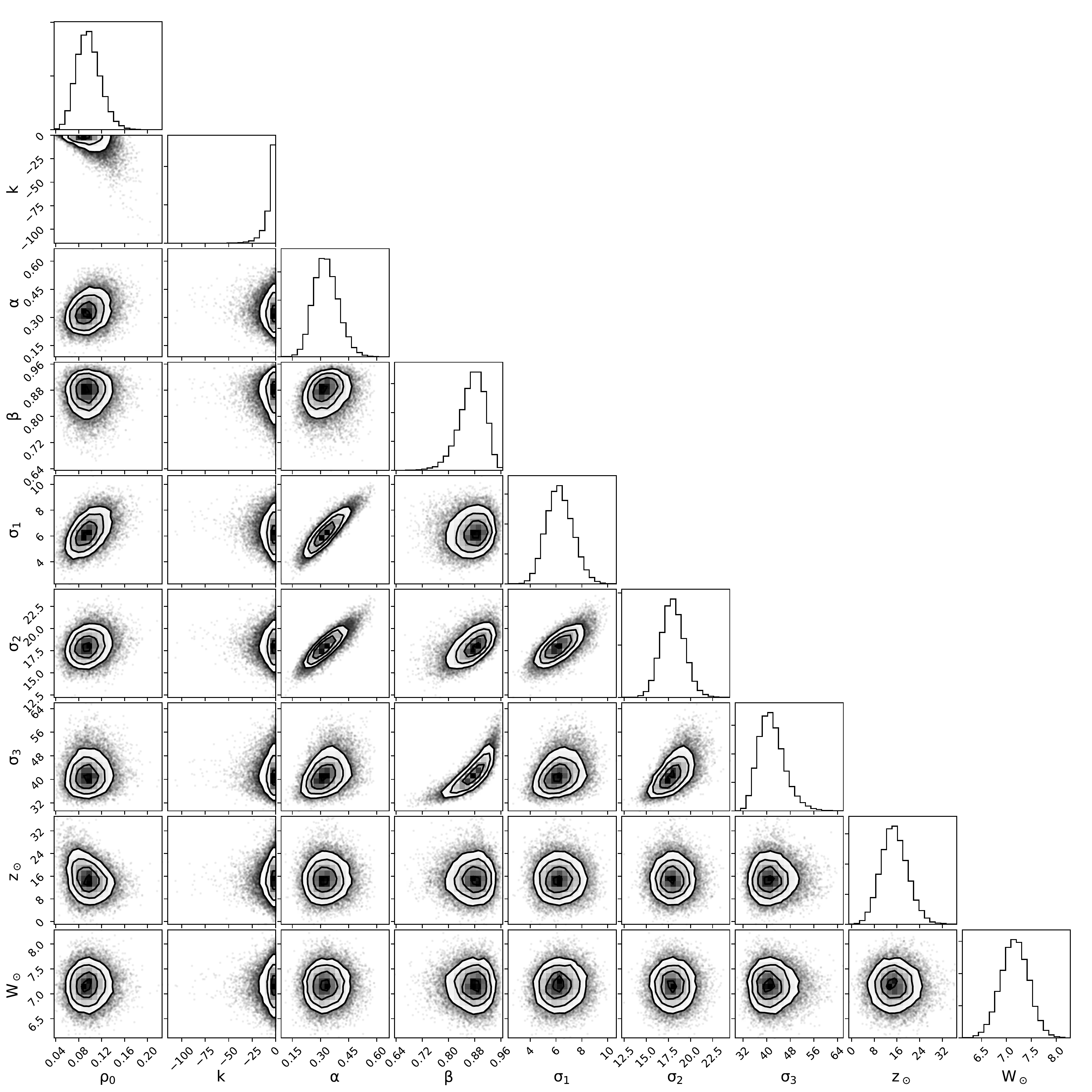}
    \caption{Population parameter posterior density of sample S2. The quantities have the following units: $\rho_0$ [$\Msun\pc^{-3}$]; $k$ [$\Msun\pc^{-3}\kpc^{-2}$]; $\alpha$ and $\beta$ [unitless]; $\sigma_{1,2,3}[\kmsec]$; $z_\odot[\pc]$; $W_\odot[\kmsec]$. The contours correspond to 0.5-, 1-, 1.5- and 2-$\sigma$ levels.}
    \label{fig:cornerS2}
\end{figure*}

\begin{figure*}
	\includegraphics[width=\linewidth]{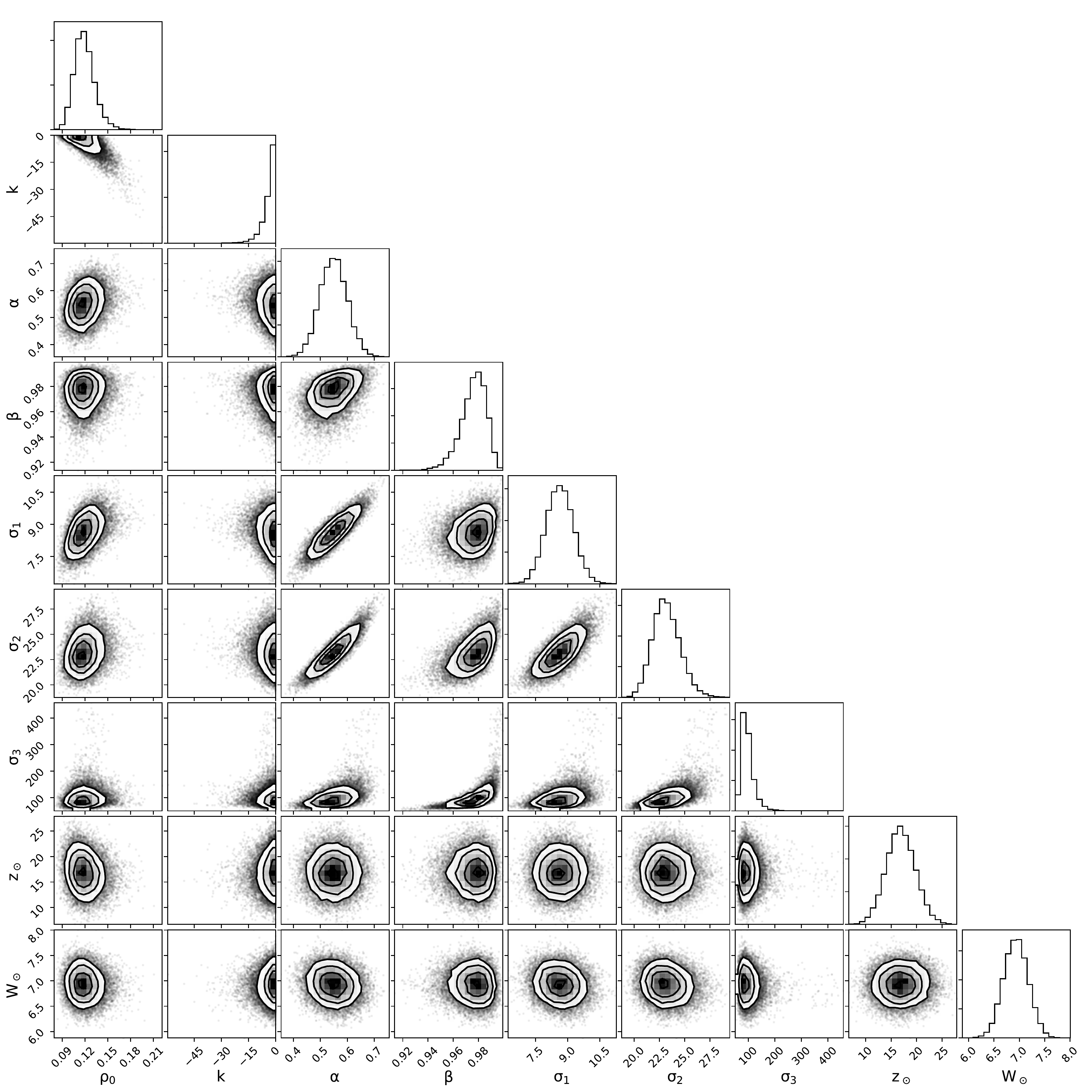}
    \caption{Population parameters posterior density of sample S3. The quantities have the following units: $\rho_0$ [$\Msun\pc^{-3}$]; $k$ [$\Msun\pc^{-3}\kpc^{-2}$]; $\alpha$ and $\beta$ [unitless]; $\sigma_{1,2,3}[\kmsec]$; $z_\odot[\pc]$; $W_\odot[\kmsec]$. The contours correspond to 0.5-, 1-, 1.5- and 2-$\sigma$ levels.}
    \label{fig:cornerS3}
\end{figure*}

\begin{figure*}
	\includegraphics[width=\linewidth]{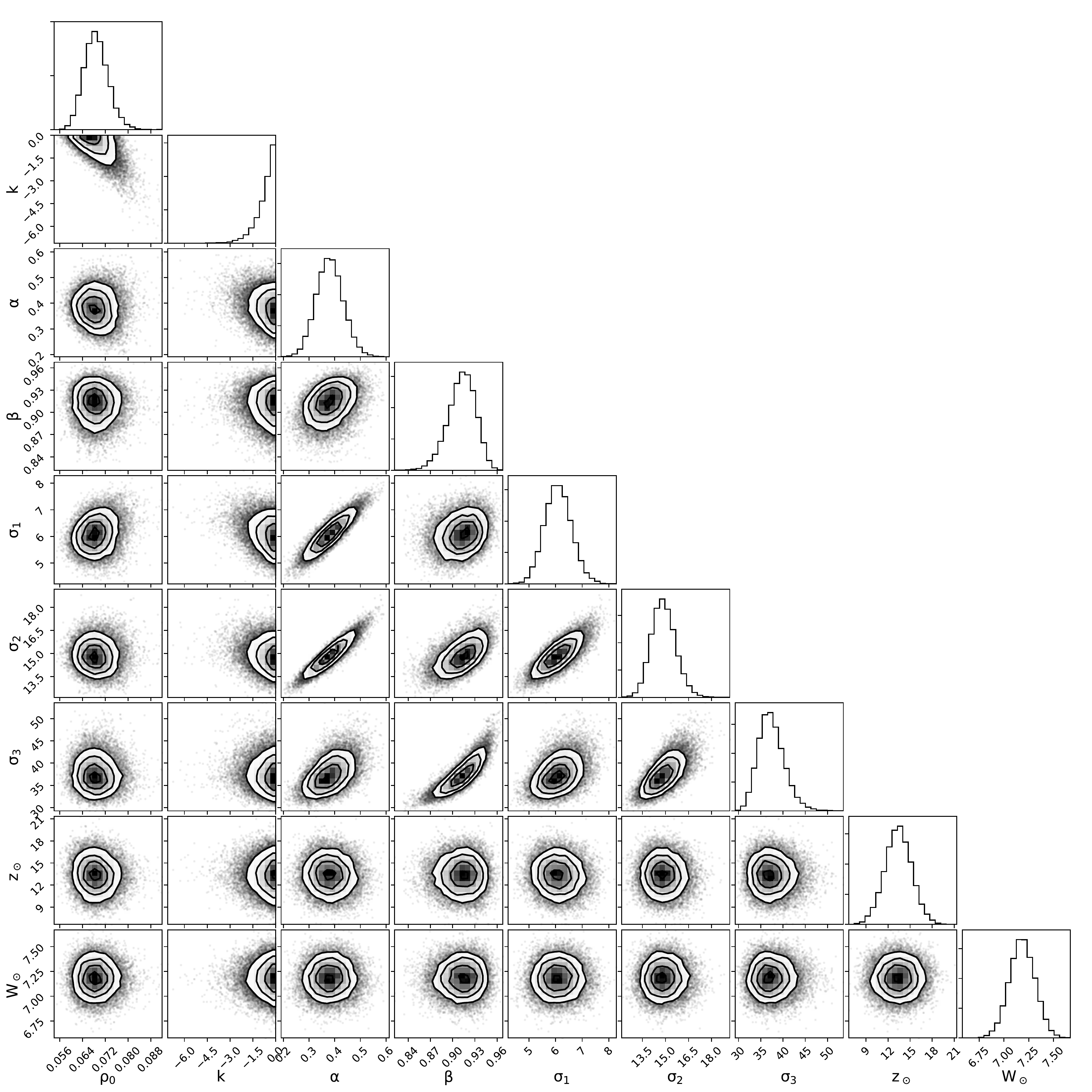}
    \caption{Population parameters posterior density of sample S4. The quantities have the following units: $\rho_0$ [$\Msun\pc^{-3}$]; $k$ [$\Msun\pc^{-3}\kpc^{-2}$]; $\alpha$ and $\beta$ [unitless]; $\sigma_{1,2,3}[\kmsec]$; $z_\odot[\pc]$; $W_\odot[\kmsec]$. The contours correspond to 0.5-, 1-, 1.5- and 2-$\sigma$ levels.}
    \label{fig:cornerS4}
\end{figure*}

The posterior densities on the population parameters of samples S1--S4, with stellar parameters marginalized, are visible in Figures \ref{fig:cornerS1}, \ref{fig:cornerS2}, \ref{fig:cornerS3}, and \ref{fig:cornerS4}. The plots are produced with the package \texttt{Corner} \citep{corner}.

%%%%%%%%%%%%%%%%%%%%%%%%%%%%%%%%%%%%%%%%%%%%%%%%%%

% Don't change these lines
\bsp	% typesetting comment
\label{lastpage}
 \end{document}